\documentclass[3p]{elsarticle}
\pdfoutput=1
\usepackage{url}
\usepackage{indentfirst}
\usepackage{gensymb}
\usepackage{graphicx}
\usepackage{pdfpages}
\usepackage{wrapfig}
\usepackage{float}
\usepackage[caption = false]{subfig}
\usepackage{amsmath,bm}
\usepackage{titlesec}
\usepackage{fancyhdr}
\usepackage{color, colortbl}
\usepackage{graphicx}
\usepackage{multirow}
\usepackage{array}
\usepackage{ragged2e}
\definecolor{Gray}{gray}{0.9}
\definecolor{White}{gray}{1}

\newcolumntype{?}{!{\vrule width 1pt}}

\graphicspath{{figures/}}
\begin{document}

\begin{frontmatter}
	
	\title{Three-Dimensional \textit{In Situ} Texture Development and Plasticity Accumulation in the Cyclic Loading of an $\alpha$-Ti Alloy}
	%% Group authors per af{}filiation:
	
	%% or include af{}filiations in footnotes:
	\author[1]{Rachel E. Lim\corref{mycorrespondingauthor}}
	\ead{relim@andrew.cmu.edu}

	\author[2]{Joel V. Bernier}
	\author[3,4]{Darren C. Pagan}
	\author[5]{Paul A. Shade}
	\author[1]{Anthony D. Rollett\corref{mycorrespondingauthor}}
	
	\ead{rollett@andrew.cmu.edu}

	\cortext[mycorrespondingauthor]{Corresponding author}

	\address[1]{Carnegie Mellon University, Pittsburgh, PA 15213, USA}
	\address[2]{Lawrence Livermore National Laboratory, Livermore, CA 94550, USA}
	\address[3]{Cornell High Energy Synchrotron Source, Ithaca, NY 14850, USA}
	\address[4]{Pennsylvania State University, University Park, PA 16802, USA}
	\address[5]{Air Force Research Laboratory, Wright-Patterson AFB, OH 45433, USA}
	
	\begin{abstract}
		 High-energy synchrotron x-rays are used to track grain rotations and the micromechanical evolution of a hexagonal Ti-7Al microstructure as it is cyclically loaded below its macroscopic yield stress. The evolution of the grains through 200 cycles reveals a continual change in von Mises stress and orientation across the entire specimen indicating a slow accumulation of plasticity even though the sample was cycled below its macroscopic yield. Grain reorientation is consistent with the development of a tension texture despite the negligible magnitude of (macroscopic) plastic strain.  It is observed that grains in a ``hard" orientation (c-axis close to the loading direction) retain more residual stress on their slip systems upon unloading than the grains in a ``soft" orientation.
	\end{abstract}
	
	
\end{frontmatter}

\section{Introduction}
	
	The accumulation of damage due to cyclic loading is known as fatigue and is a primary source of failure in structural components. Damage accumulation is sensitive to local slip within grains, making it necessary to study fatigue on the mesoscopic level \cite{DangVan1999,Constantinescu2003, Charkaluk2009}. However, in spite of recent advances in materials characterization techniques, the evolution of stress in a sample subjected to cyclic loading has not been extensively investigated experimentally on the grain scale.
	
	Previous work on the mesoscale evolution of cyclically loaded titanium has primarily been done by modeling using statistically representative microstructures \cite{Mayeur2004,Zhang2007,Smith2018} or through experiments on single crystals \cite{Tan1998,Tan1998part2,Xiao2003}. Neutron dif{}fraction experiments performed on 316 stainless steel and other cubic materials have shown the crystallographic orientation dependence in the development of intergranular residual strain \cite{Clausen1998,Wang2003,Zheng2013}. Synchrotron x-ray dif{}fraction work in this area has been focused on \textit{in situ} heterogeneous cyclic plasticity \cite{Obstalecki2014,Wong2015, Carson2017}, fatigue crack initiation and growth \cite{Herbig2011,Oddershede2012,Proudhon2017, Naragani2017, Rovinelli2018,Naragani2019}, or post mortem characterization of fatigue failure \cite{Spear2014}. However, fatigue can occur below the macroscopic yield stress, indicating that plasticity must be occurring in smaller amounts on lower length scales. Thus, this work focuses on the novel approach of using synchrotron x-rays to track the micromechanical and texture evolution of an experimental 3D microstructure with a hexagonal crystal structure being cyclically loaded below its macroscopic yield stress.

	High energy x-ray dif{}fraction microscopy (HEDM) is a synchrotron x-ray, rotating crystal technique used for micromechanical and microstructural characterization of materials in three dimensions. One of the major advantages to using this technique is its non-destructive nature which enables us to perform \textit{in situ} experiments to study material evolution. In particular, far-field high energy x-ray dif{}fraction microscopy (f{}f-HEDM) can be used to measure the grain-averaged elastic strain tensor, grain averaged-orientation, and center of mass for each of the grains in each state of a specimen \cite{Bernier2011,Oddershede2010}. More specifics on this technique can be found in Bernier et al. \cite{Bernier2020}. Significant previous studies utilizing the HEDM technique have been reported on the Ti-7Al material being studied here \cite{Lienert2009,Schuren2015,Chatterjee2016,Beaudoin2017, Chatterjee2017, Turner2017,Pagan2017, Pagan2018,Chatterjee2019}. Ti-7Al a model alloy for the hexagonal $\alpha$-phase of the titanium alloy Ti-6Al-4V which is commonly used in aerospace and biomedical applications. The combination of a relatively high flow strength and relatively compliant elastic response lead to large elastic strains that are easily observable with dif{}fraction; that, combined with the ability to process a microstructure with large, pristine grains make Ti-7Al an ideal material for f{}f-HEDM studies.

	\begin{figure}[b!]
		\centering
		\subfloat[Prismatic $\langle$a$\rangle$]{\includegraphics[scale=0.05]{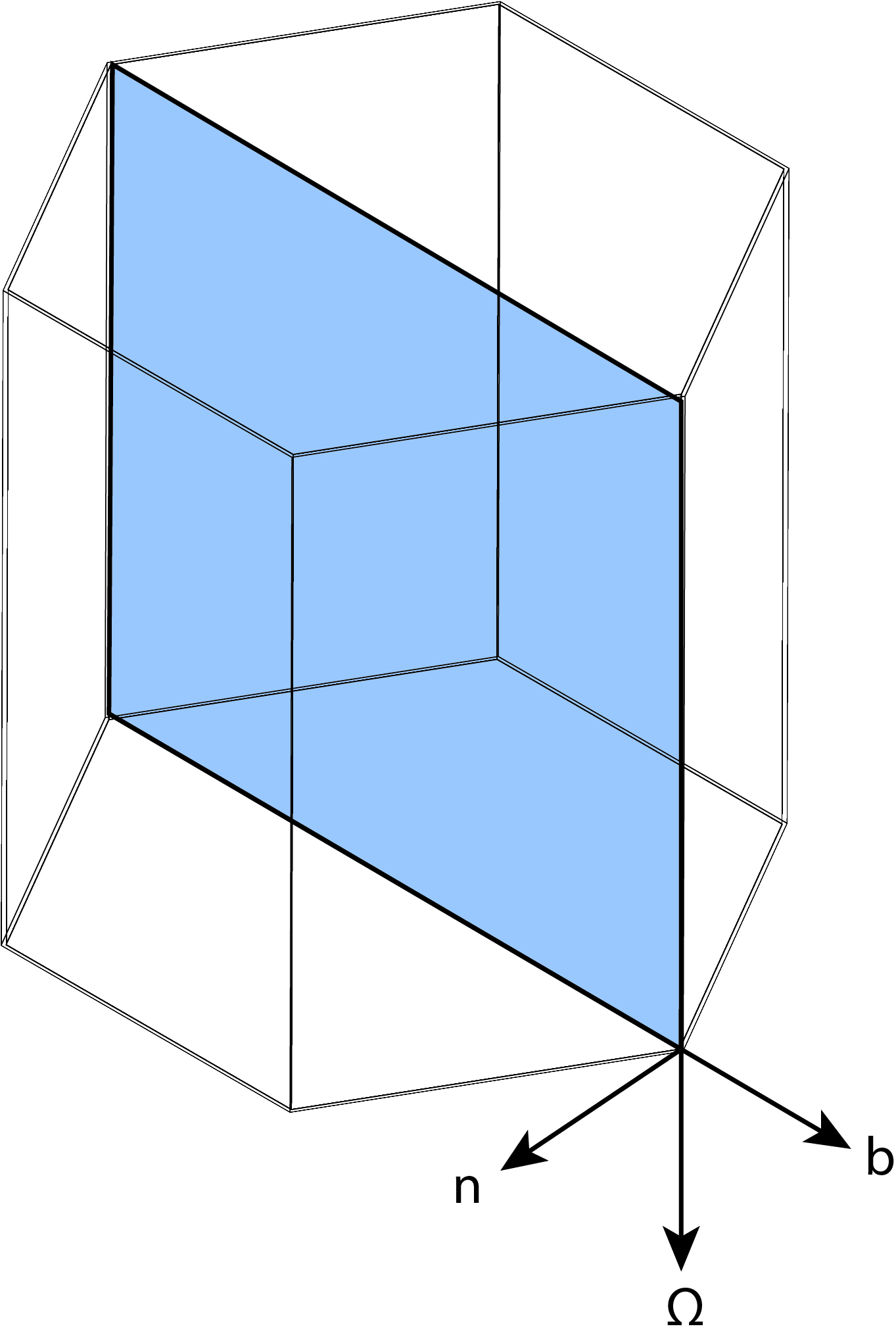}} \hspace{15pt}
		\subfloat[Basal $\langle$a$\rangle$]{\includegraphics[scale=0.05]{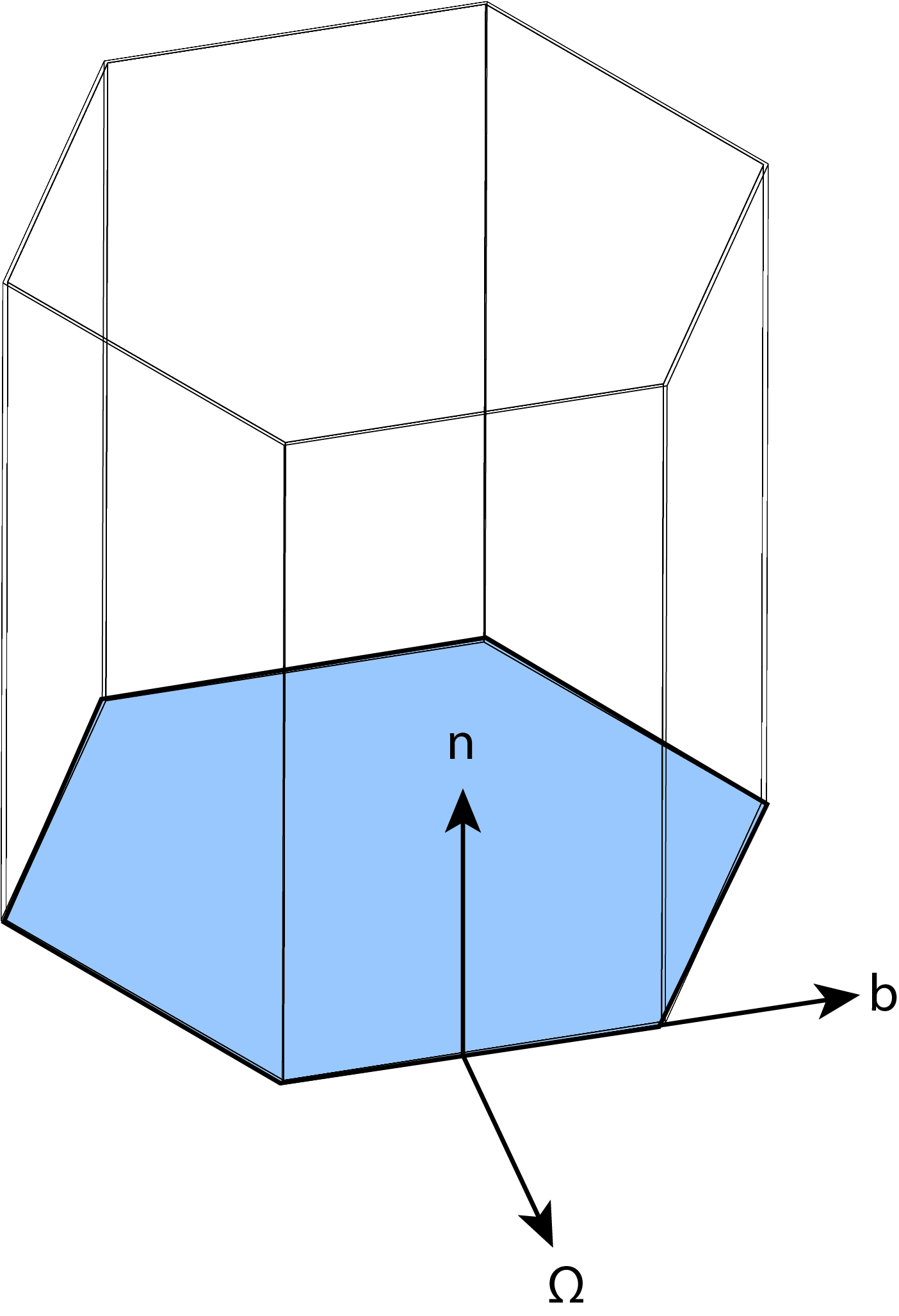}} \hspace{15pt}
		%\subfloat[Basal $\langle$a\textsubscript{1}+a\textsubscript{2}$\rangle$]{\includegraphics[scale=0.06]{basal2.jpg}} \\
		\subfloat[Pyramidal $\langle$a$\rangle$]{\includegraphics[scale=0.05]{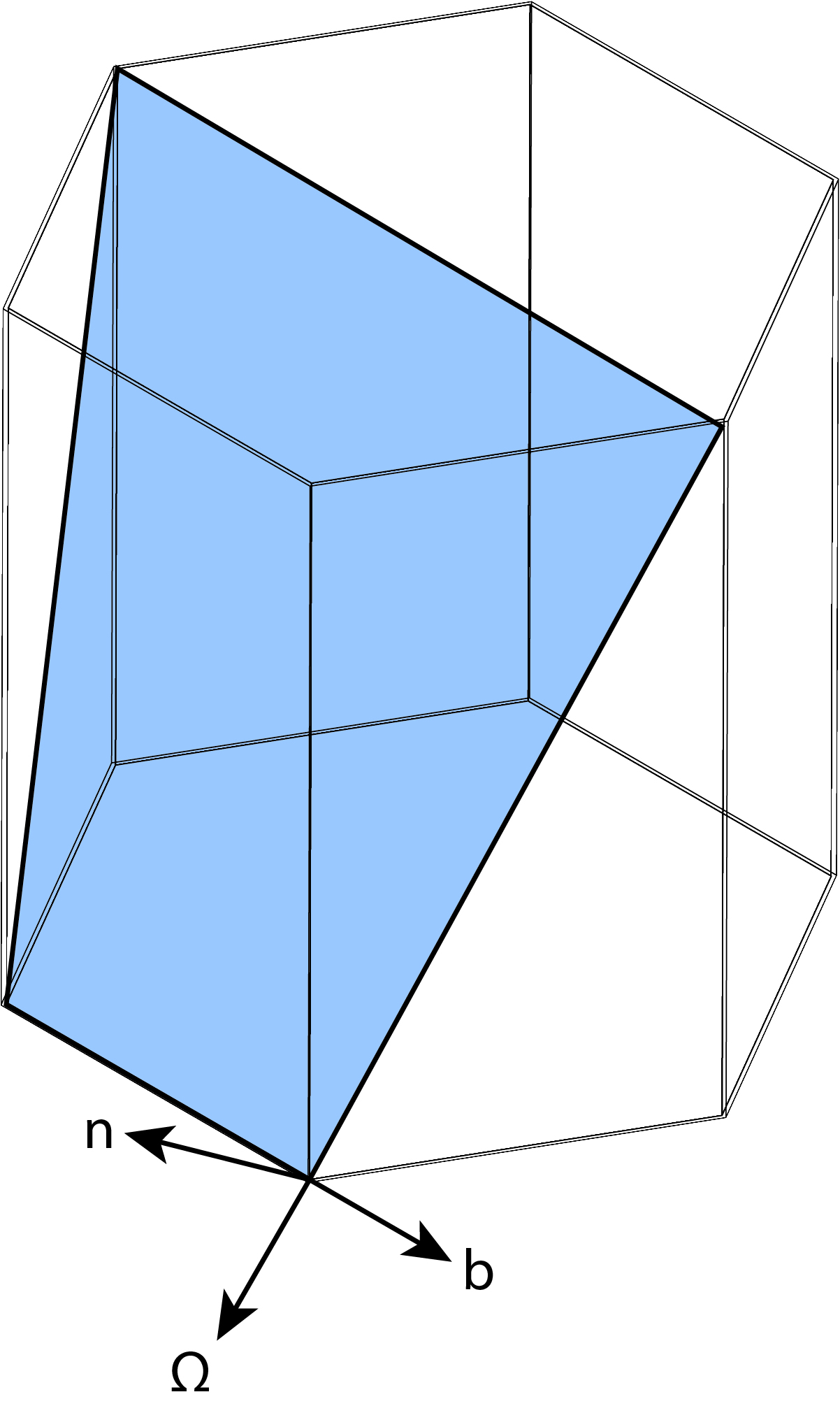}} \\
		\subfloat[1st Order \newline Pyramidal $\langle$c+a$\rangle$]{\includegraphics[scale=0.05]{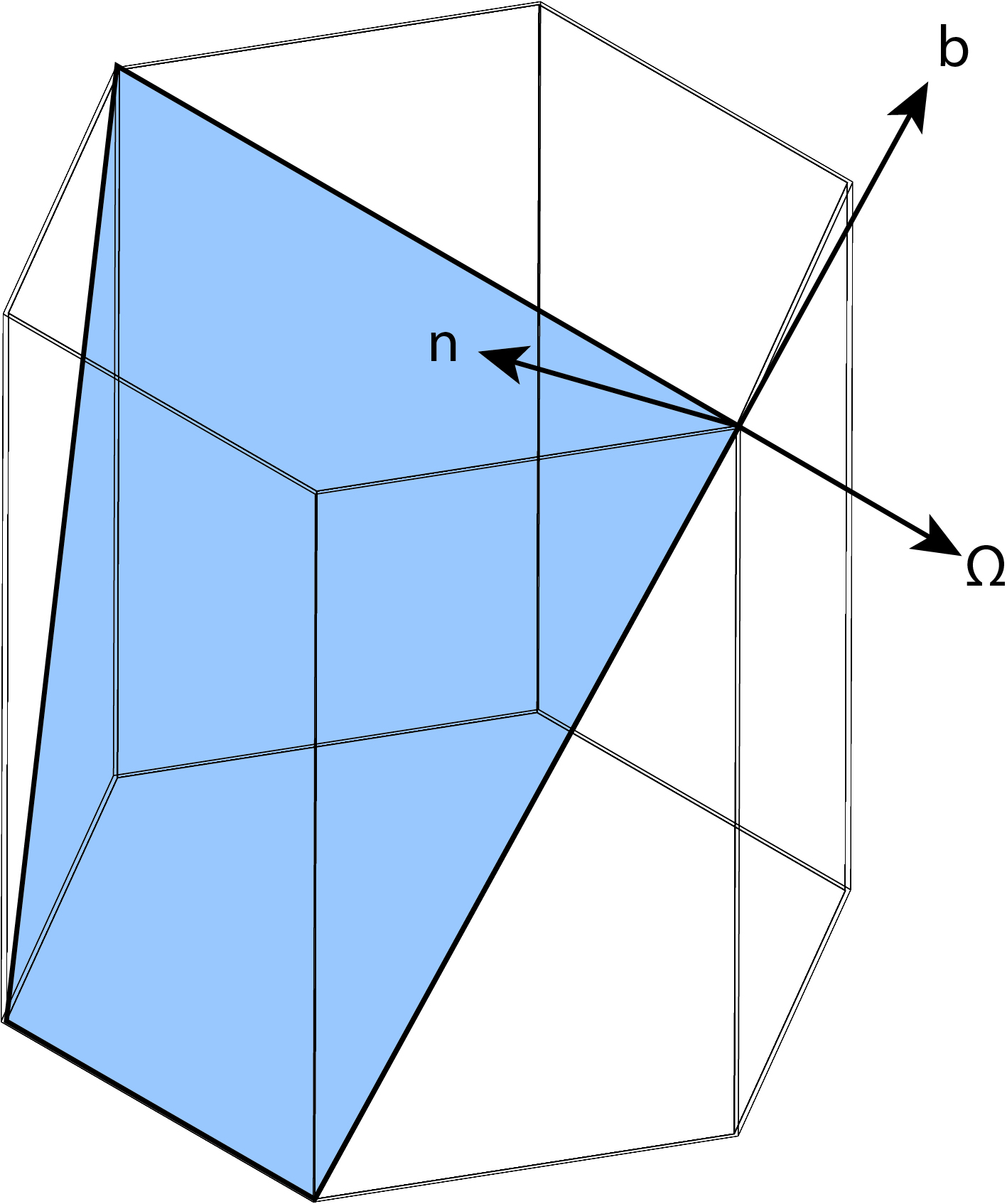}} \hspace{15pt}
		\subfloat[2nd Order \newline Pyramidal $\langle$c+a$\rangle$]{\includegraphics[scale=0.05]{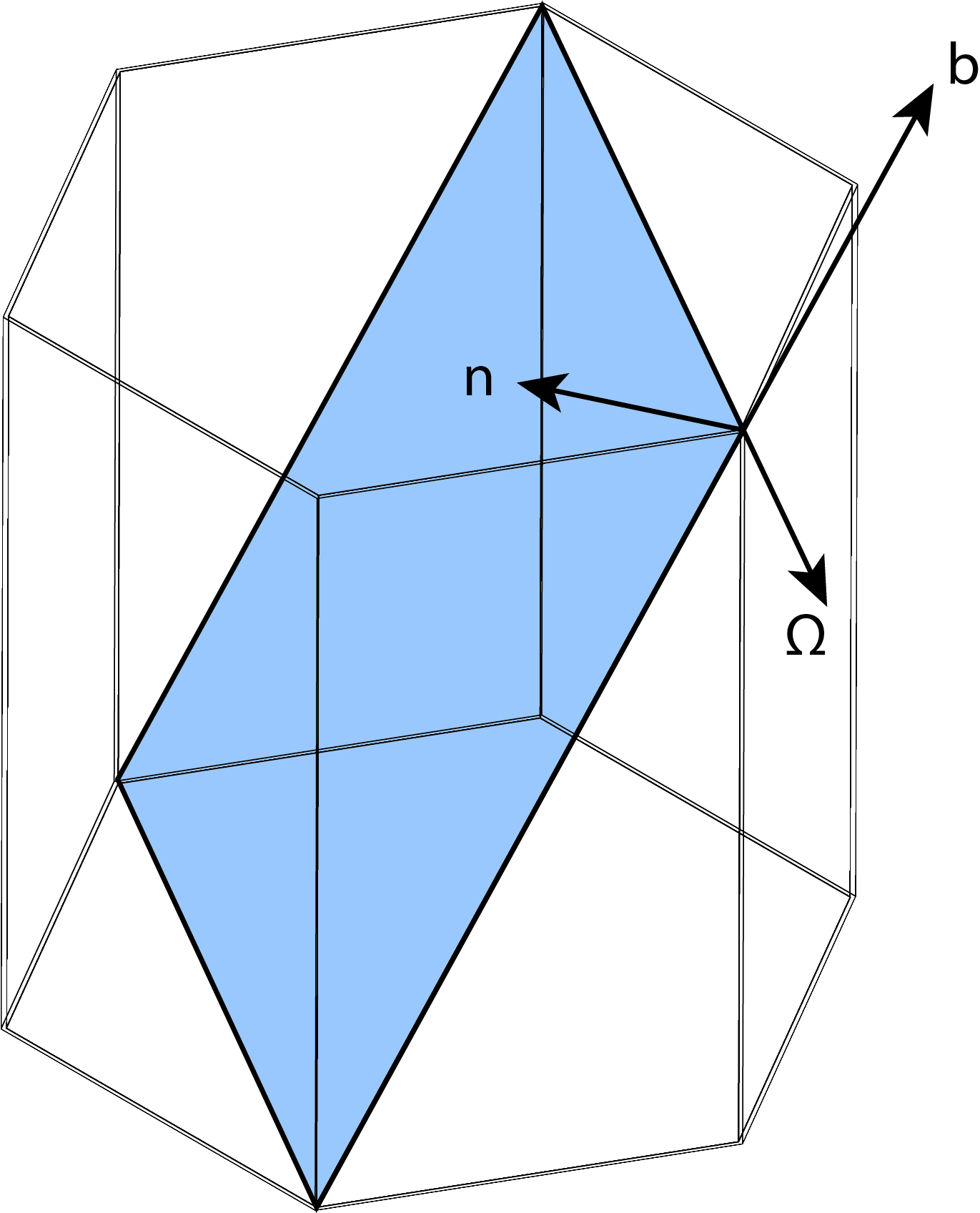}}
		\caption{The five major slip systems for the hexagonal crystal system marked with the slip plane normal, \textbf{n}, slip direction, \textbf{b}, and reorientation axis, $\Omega$.}
		\label{fig:slip systems}
	\end{figure}

	Plastic deformation in Ti-7Al is accommodated on slip systems of varying strengths which are denoted in Figure \ref{fig:slip systems} and Table \ref{tab:slip systems} \cite{Williams1968,Tan1998, Williams2002,Lutjering2007,Lienert2009,Pagan2017,Tong2018}. As seen in Table~\ref{tab:slip systems}, the prismatic $\langle$a$\rangle$ system has the lowest critical resolved shear stress (CRSS) followed by the basal $\langle$a$\rangle$ system then the pyramidal $\langle$a$\rangle$ system. Thus, grains oriented to activate $\langle$a$\rangle$ slip will preferentially slip on the prismatic $\langle$a$\rangle$ systems. In order to close the yield surface, slip may also occur on the pyramidal $\langle$c+a$\rangle$ systems. When slip occurs on a single slip system, the axis about which the grain rotates, \textbf{$\Omega$}, can be calculated using \textbf{b}$\times$\textbf{n} where  \textbf{b} is the slip direction, and \textbf{n} is the slip plane normal (Table \ref{tab:slip systems}). Consequently, this reorientation axis can indicate which slip system is active when a grain's change in orientation is known. However, this only works for the case where a single slip system is being activated in a grain. When multiple slip occurs, the axis is less useful because it becomes more dif{}ficult to separate out the ef{}fects of individual slip systems without knowing their separate contributions. Furthermore, the addition of $>5$ wt$\%$ of aluminum to titanium encourages the formation of short range order and $\alpha_2$ Ti\textsubscript{3}Al nanoprecipitates, which have previously been observed in Ti-7Al \cite{Venkataraman2017}. In turn, these suppress twinning and give the  slip systems their strength \cite{Neeraj2001,Yoo2002,VandeWalle2002,Fitzner2016, Venkataraman2017, Gardner2020}. Subsequent shearing of the $\alpha_2$ from dislocation movement causes these slip systems to soften \cite{Pagan2017}.
	
	In this work, f{}f-HEDM is used to measure the evolution of the micromechanical and microstructural state of grains, individually and as an ensemble, in a sample of Ti-7Al through 200 loading cycles. %Both the general texture development of the ensemble of grains and the rotations of individual grains are examined.

	\begin{table}
		\centering
		\begin{tabular}{c|c|c|c|c}
			
			\textbf{Slip System}  & \textbf{b} & \textbf{n} & \textbf{$\Omega$} & \textbf{CRSS} (MPa)  \cite{Pagan2017}  \\ \cline{1-5}
			Prismatic $\langle$a$\rangle$  &  $\langle1\bar{2}10\rangle$ & $\{10\bar{1}0\}$ & $\langle0001\rangle$ & 248  \\
			Basal $\langle$a$\rangle$  &  $\langle1\bar{2}10\rangle$ & $\{0001\}$ &  $\langle10\bar{1}0\rangle$ & 253 \\
			%Basal $\langle$a\textsubscript{1}+a\textsubscript{2}$\rangle$  &  $\langle1\bar{1}00\rangle$ & $\{0001\}$ &  $\langle2\bar{1}\bar{1}0\rangle$ & \\
			Pyramidal $\langle$a$\rangle$  &  $\langle11\bar{2}0\rangle$  & $\{1\bar{1}01\}$ &  $\langle01\bar{1}2\rangle$ & 255 \\
			1st Order Pyramidal $\langle$c+a$\rangle$   &  $\langle\bar{1}\bar{1}2\bar{3}\rangle$ &  $\{1\bar{1}01\}$   & $\langle13\hspace{2pt}\bar{8}\hspace{1pt}\bar{5}\hspace{1pt}\bar{3}\rangle$ & 270 \\
			2nd Order Pyramidal $\langle$c+a$\rangle$  &   $\langle\bar{1}\bar{1}2\bar{3}\rangle$ &  $\{11\bar{2}2\}$  & $\langle1\bar{1}00\rangle$ &
		\end{tabular}
	\caption{The family of slip directions \textbf{b}, family of slip plane normals, \textbf{n}, and familes of reorientation axes, \textbf{$\Omega$} and the critical resolved shear stress (CRSS) for the five major slip systems for the hexagonal crystal system.}
	\label{tab:slip systems}
	\end{table}

\section{Experiment}

\begin{figure}[t!]
		\centering
		\subfloat[]{\includegraphics[height=0.2\textheight]{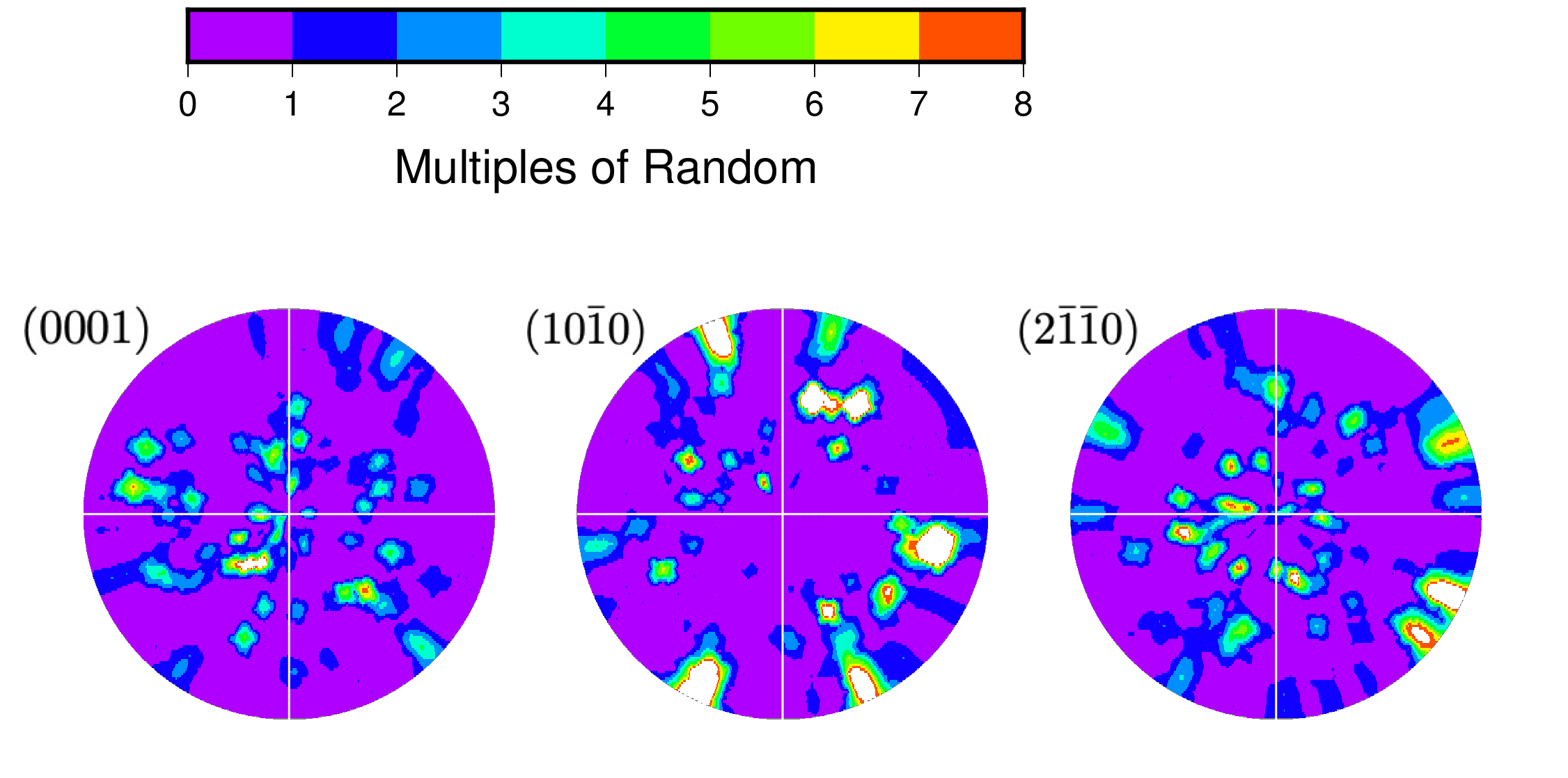}} \\
		\subfloat[]{\includegraphics[height=0.2\textheight]{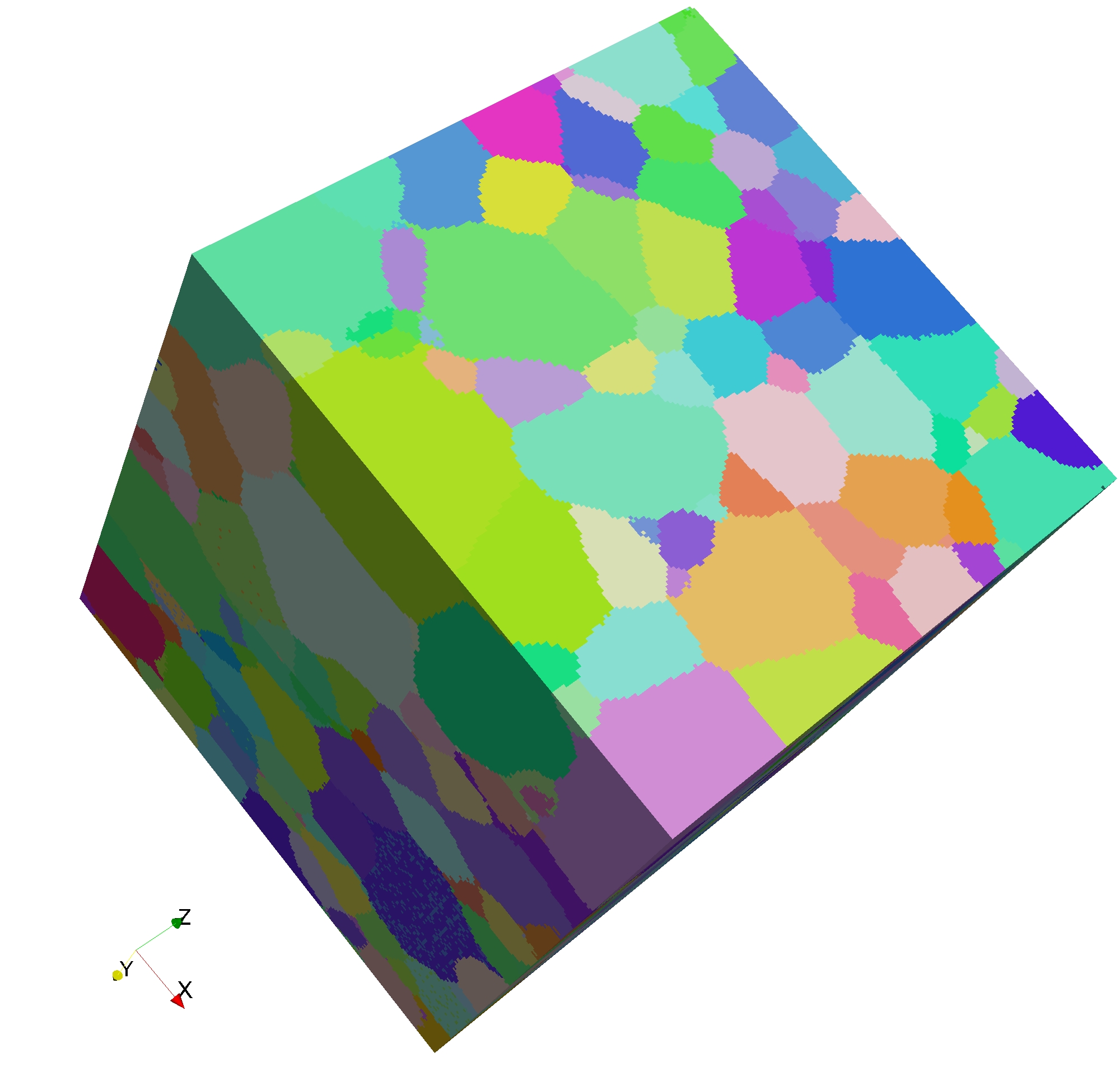} \includegraphics[height=0.05\textheight]{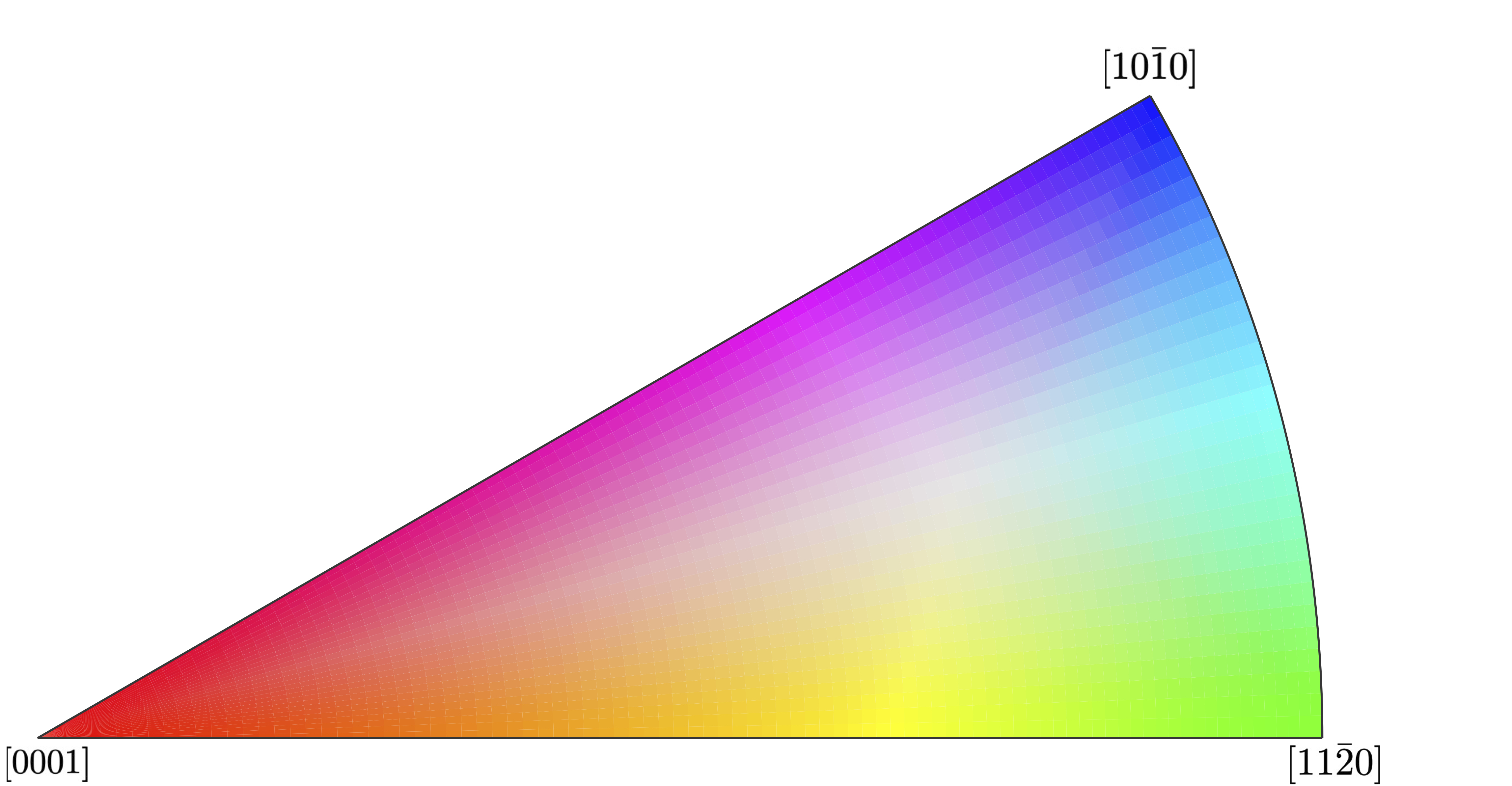}}
		\caption{(a) Contour pole figures showing that the c-axes lie mostly perpendicular to the tensile axis. (b) The nf-HEDM volume (IPF coloring with respect to the loading direction) was reconstructed using HEXRD from the grain-averaged orientations found in the f{}f-HEDM data. } 
		\label{fig:NF_volume}	
	\end{figure}

%\subsection{Sample}
	The Ti-7Al material for this experiment was cast as an ingot, hot isostatic pressed to eliminate porosity, extruded, heat treated at 962 \degree C for 24 hours, then air cooled to form large equiaxed grains \cite{Pilchak2013}. A sample was subsequently electrically discharged machined (EDM) from the bulk material, which left behind a notable amount of grain scale residual strains around the surface of the gauge section as shown in Figure \ref{fig:strain22_probplot}c. The microstructure of the region of interest in this work is shown in Figure \ref{fig:NF_volume}b with accompanying pole figures to detail the texture of the sample is in Figure \ref{fig:NF_volume}a which shows that the c-axes lie predominantly perpendicular to the tensile axis.

%\subsection{Experiment Description}
	\begin{figure}[t]
		\centering
		\includegraphics[width=0.6\textwidth]{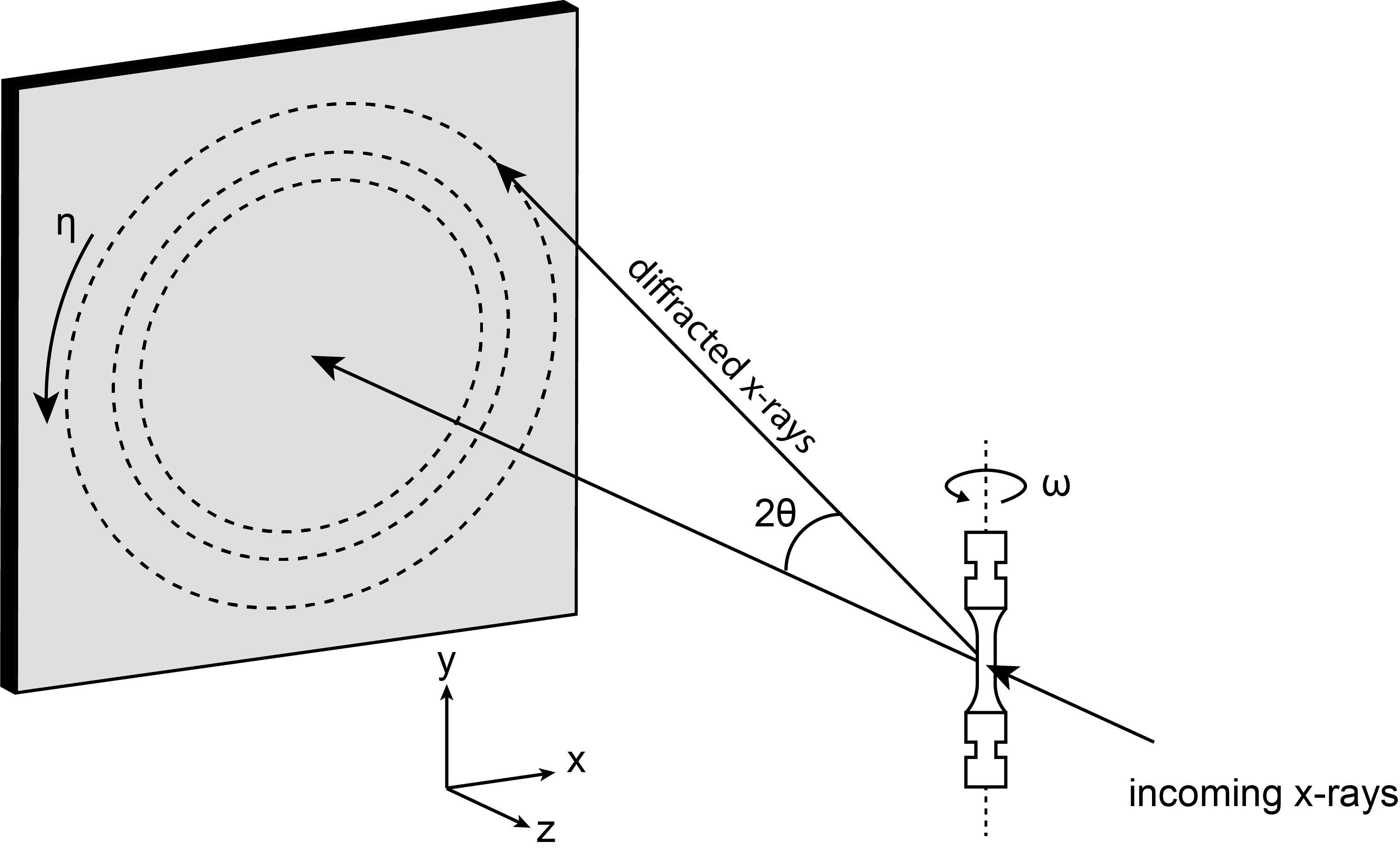}
		\caption{Schematic of the f{}f-HEDM experimental setup at CHESS marked where $\omega$ is the rotation angle of the sample, and $\eta$ is the azimuthal coordinate on the detector, and $2\theta$ is the dif{}fraction angle.}
		\label{fig:HEDM_setup}
	\end{figure}

\begin{figure}[t]
		\subfloat[]{\includegraphics[height=0.27\textwidth]{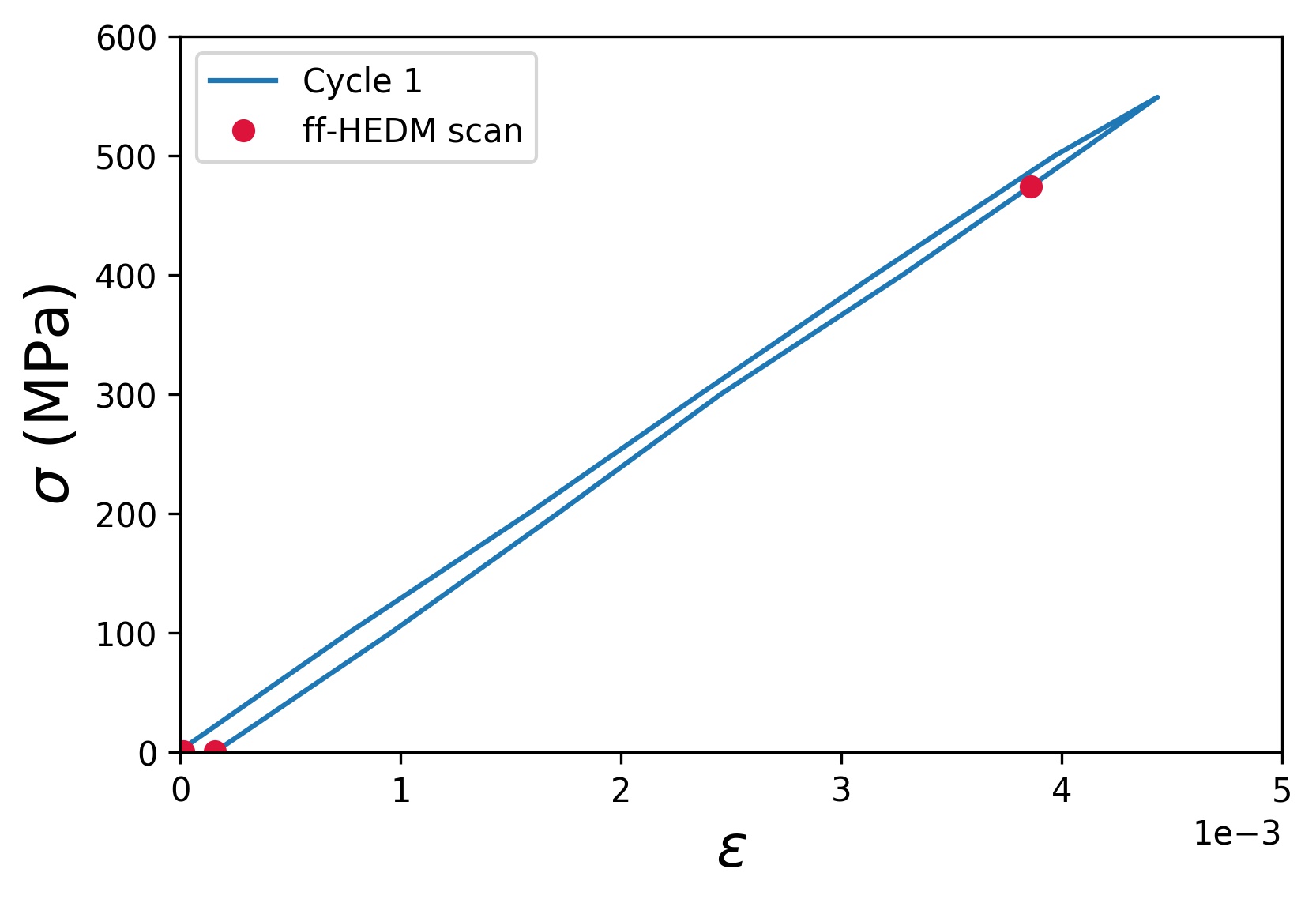}}\hspace{5pt}
		\subfloat[]{\includegraphics[height=0.27\textwidth]{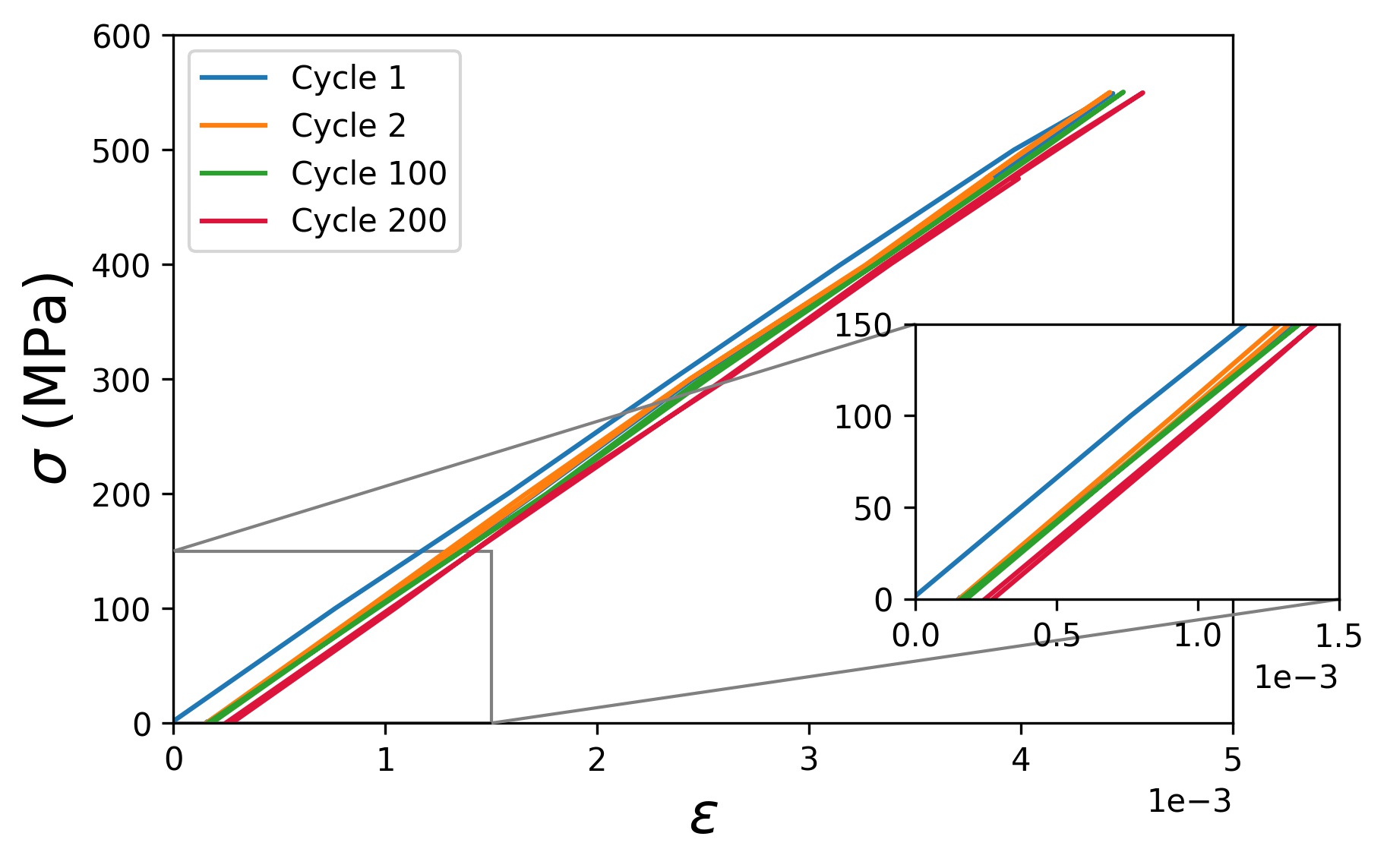}}
		\centering
		\caption{(a) The cyclic macroscopic stress-strain curves was calculated using DIC for the first load cycle with the points marked where f{}f-HEDM scans were taken. (b) The cyclic stress-strain curves for the 1st, 2nd, 100th, and 200th cycles are shown with some plasticity occurring during the first cycle and the other cycles showing a mostly elastic response.} 
		\label{fig:ti7_stress_strain}	
	\end{figure}

	Tensile cycles (R=0) to 585 MPa ($\sim$90~\% of yield) were performed in load control on a sample of Ti-7Al using the RAMS2 load frame \cite{Shade2015} at the F2 beamline at the Cornell High Energy Synchrotron Source (CHESS) with the geometry shown in Figure \ref{fig:HEDM_setup}. The sample had an 8mm gauge length and a 1 mm x 1 mm cross-sectional area. Near-field and far-field HEDM scans were taken in the initial unloaded state. Then, the sample was loaded to 585 MPa and unloaded to 505 MPa to minimize stress relaxation during f{}f-HEDM measurements, and another f{}f-HEDM scan was taken over the same approximate volume. The sample was unloaded back to 0 N, and a final f{}f-HEDM scan was taken (Fig. \ref{fig:ti7_stress_strain}a). This process was repeated at cycles 2, 5, 10, 20, 30, and every 10 cycles until a total of 200 cycles was reached. 
	
	The f{}f-HEDM scans were taken for a full 360$\degree$ sweep at an $\omega$ interval of 0.25$\degree$ using an energy of 61.3 keV. For these scans, data was collected for a 1 mm tall volume in six 60$\degree$ wedges using a GE amorphous silicon area detector (2048 x 2048 pixels, 200 $\mu$m pixel size). Reduction of the f{}f-HEDM data was done using the HEXRD package (\url{https://github.com/HEXRD/hexrd}) \cite{Bernier2011}. The dif{}fraction was fit to obtain the grain averaged orientation ($\mathbf{g}$), center of mass, and elastic strain tensor ($\bm{\varepsilon}$). The reduced data was filtered with a threshold of normalized sum of square residuals $< 2\times 10^{-3}$ which left 502 of the original 773 grains to be used for analysis \cite{Bernier2020}. Then, the stress tensor, $\bm{\sigma}$, was calculated from the single crystal stif{}fness tensor, $\mathbf{C}$, \cite{Fisher1964,Pagan2017} and the elastic strain tensor for each of the grains in each scan using $\bm{\sigma}_{ij}= \mathbf{C}_{ijkl}\bm{\varepsilon}_{kl}$.

% a = 2.924942
% c = 4.673149

\section{Results}

	For the macroscopic stress-strain response shown in Figure \ref{fig:ti7_stress_strain}b, the macroscopic strain was measured using digital image correlation performed on optical images of the sample surface and macroscopic stress was calculated from a load cell placed above the specimen. The unloading curve for the first cycle has shifted slightly from the loading for that cycle showing that some plasticity occurred, while the curve for the second cycle exhibits a stable elastic response that lies on top of the unloading of the first cycle. By the $100^{th}$ cycle, the curve has shifted just slightly, and the 200th cycle has shifted slightly more showing small amounts of plasticity occurring over the course of each 100 cycles.
	
\subsection{First Cycle}

\begin{figure}[t!]
		\centering
		\subfloat[]{\includegraphics[height=0.27\textwidth]{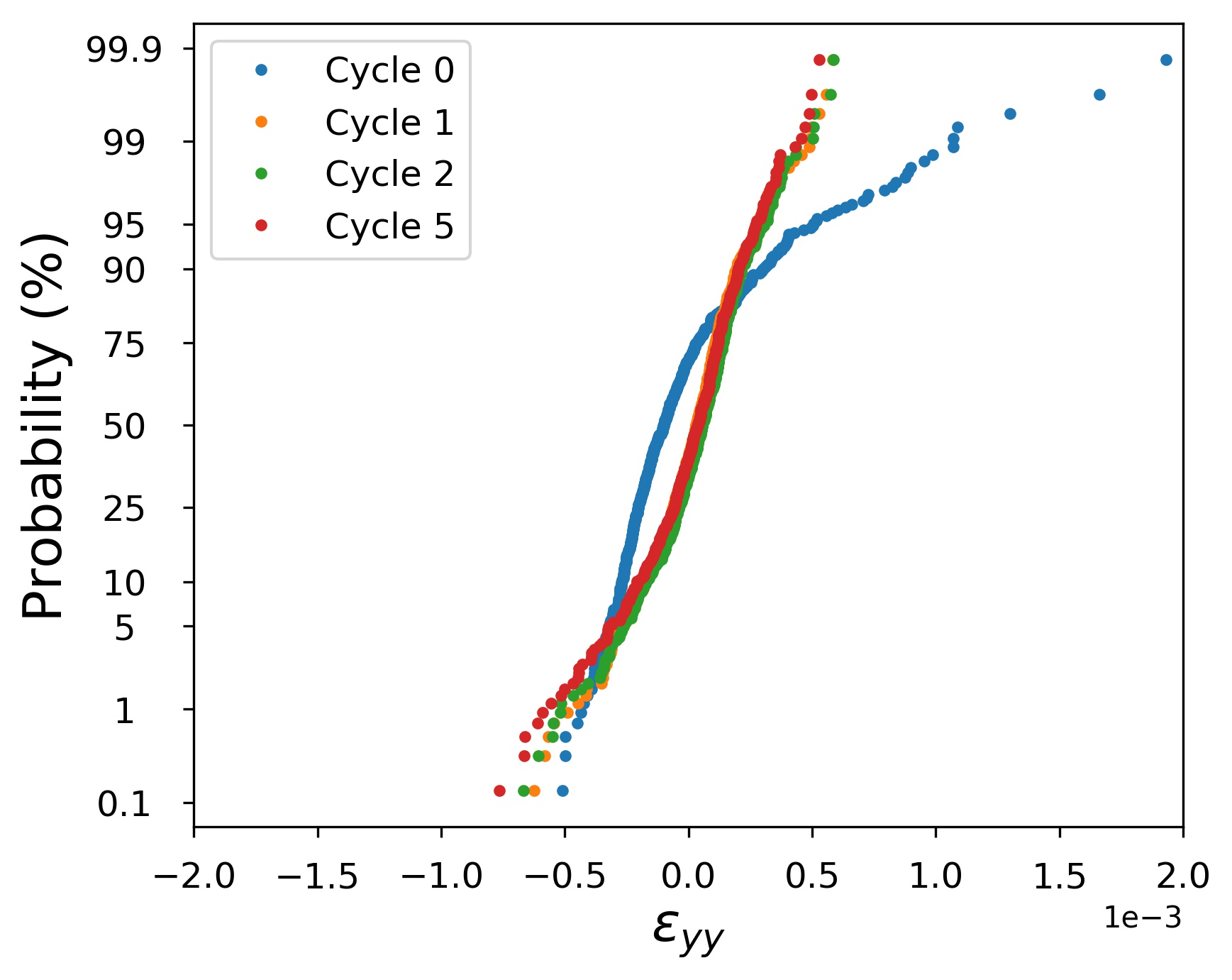}} \hspace{5pt}
		\subfloat[]{\includegraphics[height=0.27\textwidth]{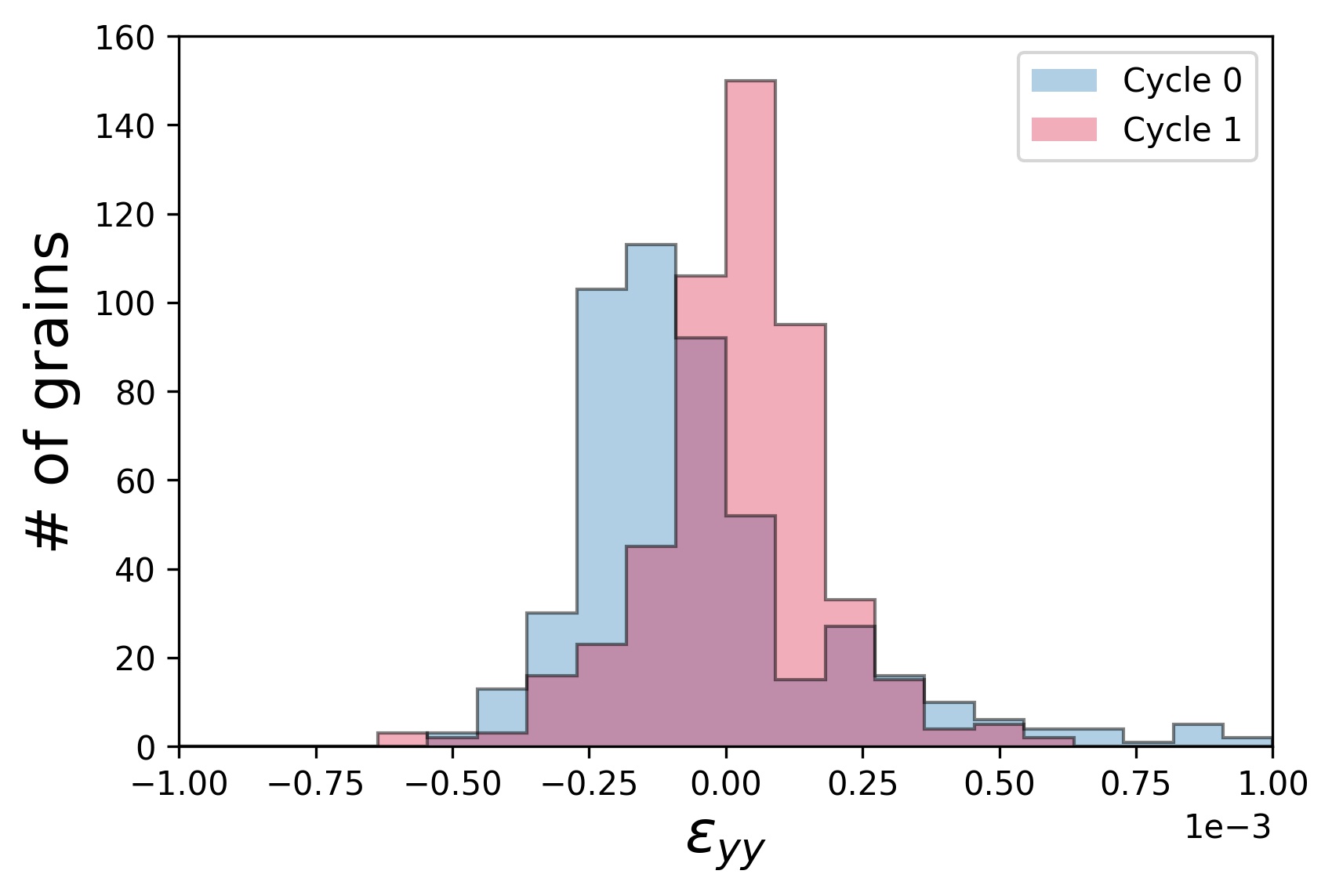}} \\
		\subfloat[]{\includegraphics[height=0.27\textwidth]{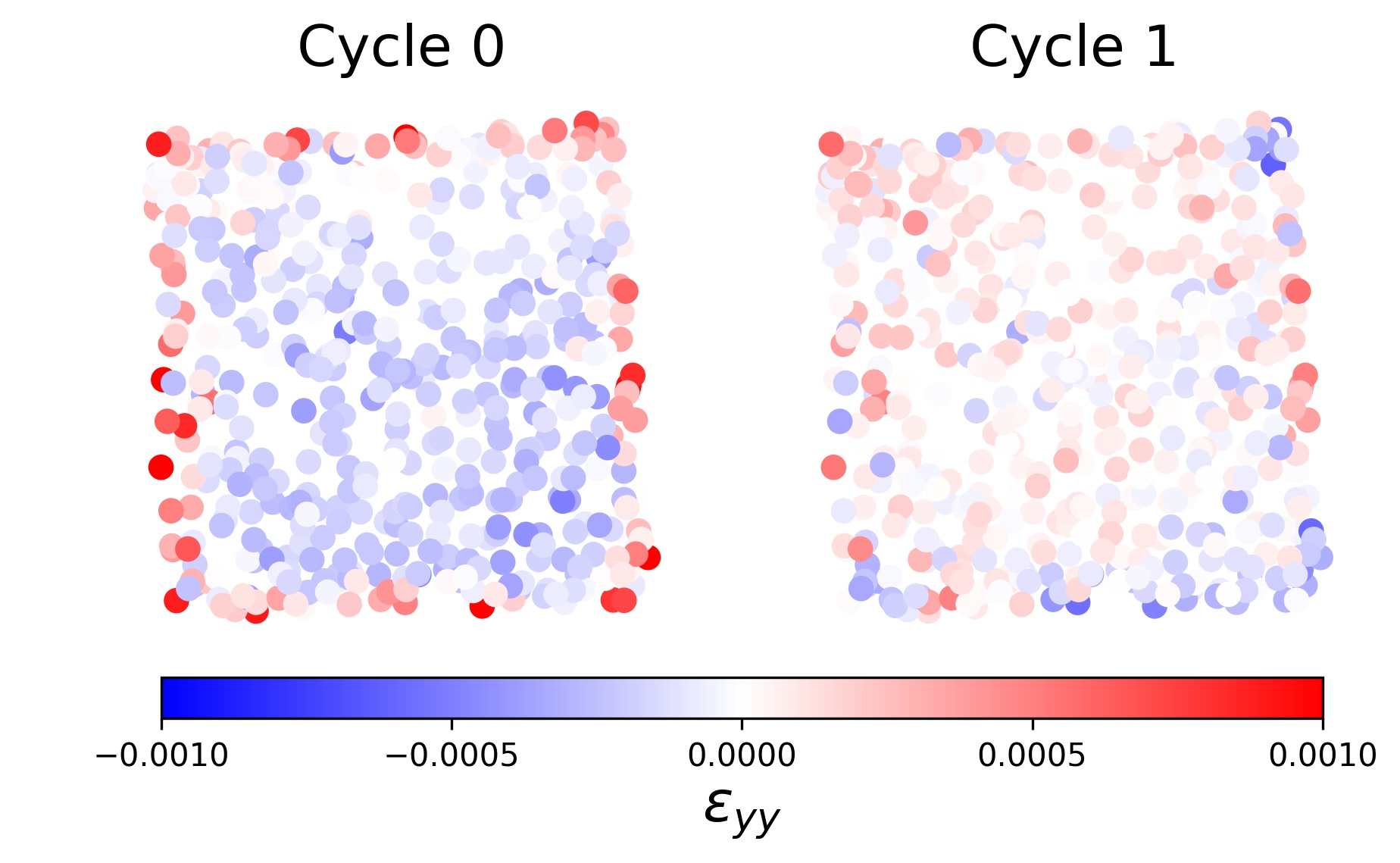}}
		\includegraphics[height=0.1\textheight]{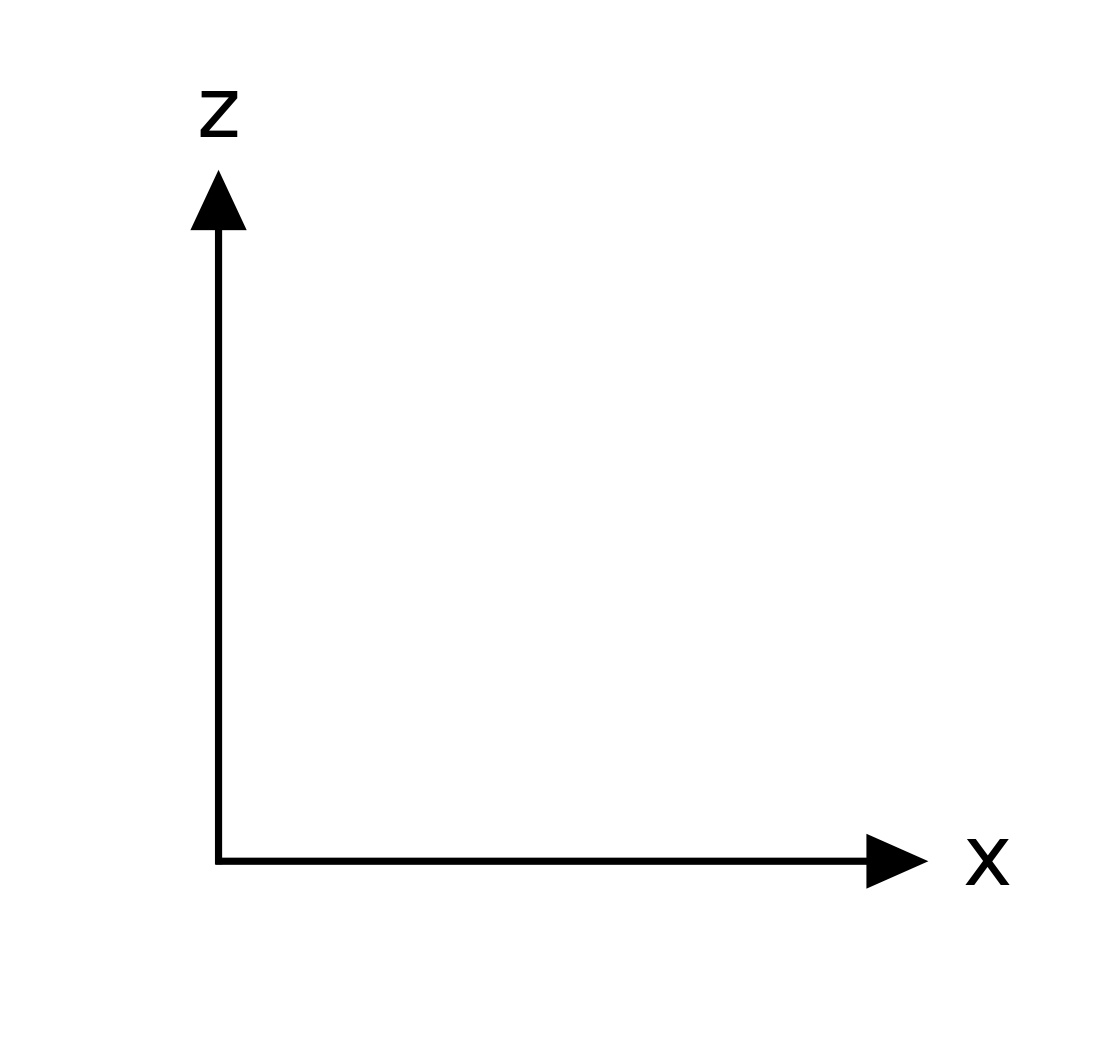}\\
		\subfloat[]{\includegraphics[height=0.27\textwidth]{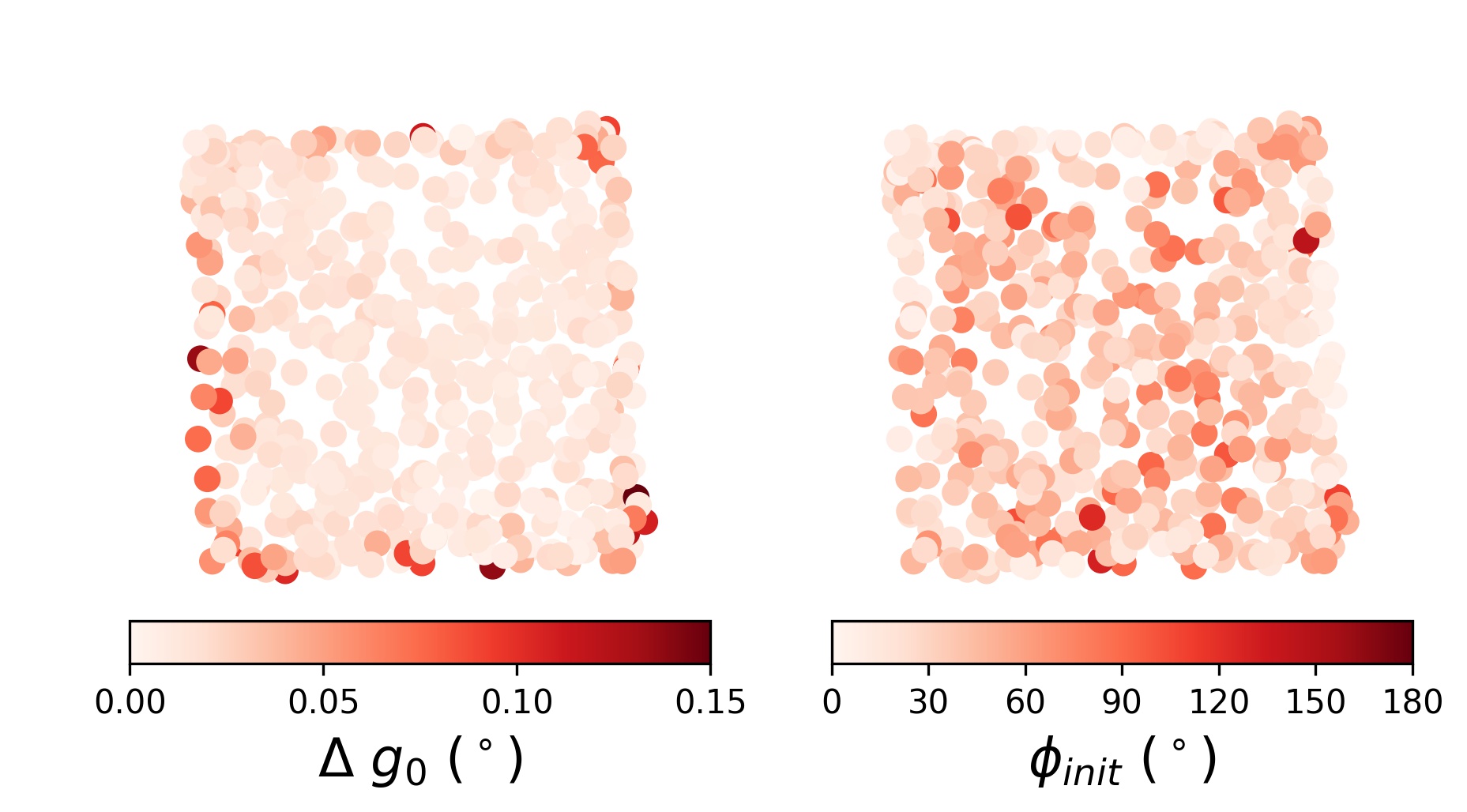}}\hspace{5pt}
%		\subfloat{\includegraphics[height=0.15\textheight]{xz.jpg}}
		\caption{(a) The normal probability distribution of $\varepsilon_{yy}$ shows a change in the distribution from the initial state to the states after 1, 2, and 5 cycles. (b) The histogram of $\varepsilon_{yy}$ shift towards 0 during the first cycle. (c) $\varepsilon_{yy}$ is plotted according to the distribution in the x-z plane with color corresponding to the magnitude of the strain. (d) The grains with the largest $\Delta g$ after the first cycle are located on the surface of the sample. The grains with the smallest $\phi_{init}$ are also on the surface of the sample. } 
		\label{fig:strain22_probplot}	
	\end{figure}

	On the grain level, the spread of the distribution of the strain in the loading direction ($\varepsilon_{yy}$) can be analyzed using a normal probability plot which compares the data to a normal distribution and would be represented by a straight line. Figure \ref{fig:strain22_probplot}a shows that while cycles 1, 2, and 5 appear to be reasonably approximated by the same distribution, the initial state is significantly dif{}ferent. The tail on the initial distribution means that a portion of the grains have an initial strain higher than expected by the bulk of the distribution. In Figure \ref{fig:strain22_probplot}b, the histograms of $\varepsilon_{yy}$ show a shift from a peak at a negative strain to a peak at zero. The spatial distribution of this change can be seen in Figure \ref{fig:strain22_probplot}c where in the initial state, grains on the outer surface have a tensile $\varepsilon_{yy}$ while the inner grains have a compressive $\varepsilon_{yy}$. Then, after the first cycle, the spatial heterogeneity of strain has been eliminated via dislocation activity in the surface grains and the strain values are more nearly randomly distributed. Further supporting the evidence of surface grains yielding during the first cycle, in Figure \ref{fig:strain22_probplot}d, the grains with the largest change in orientation are on the surface of the sample.
	
\subsection{Evolution of Measured Quantities}

\begin{figure}[t]
		\centering
		{\includegraphics[width=0.5\textwidth]{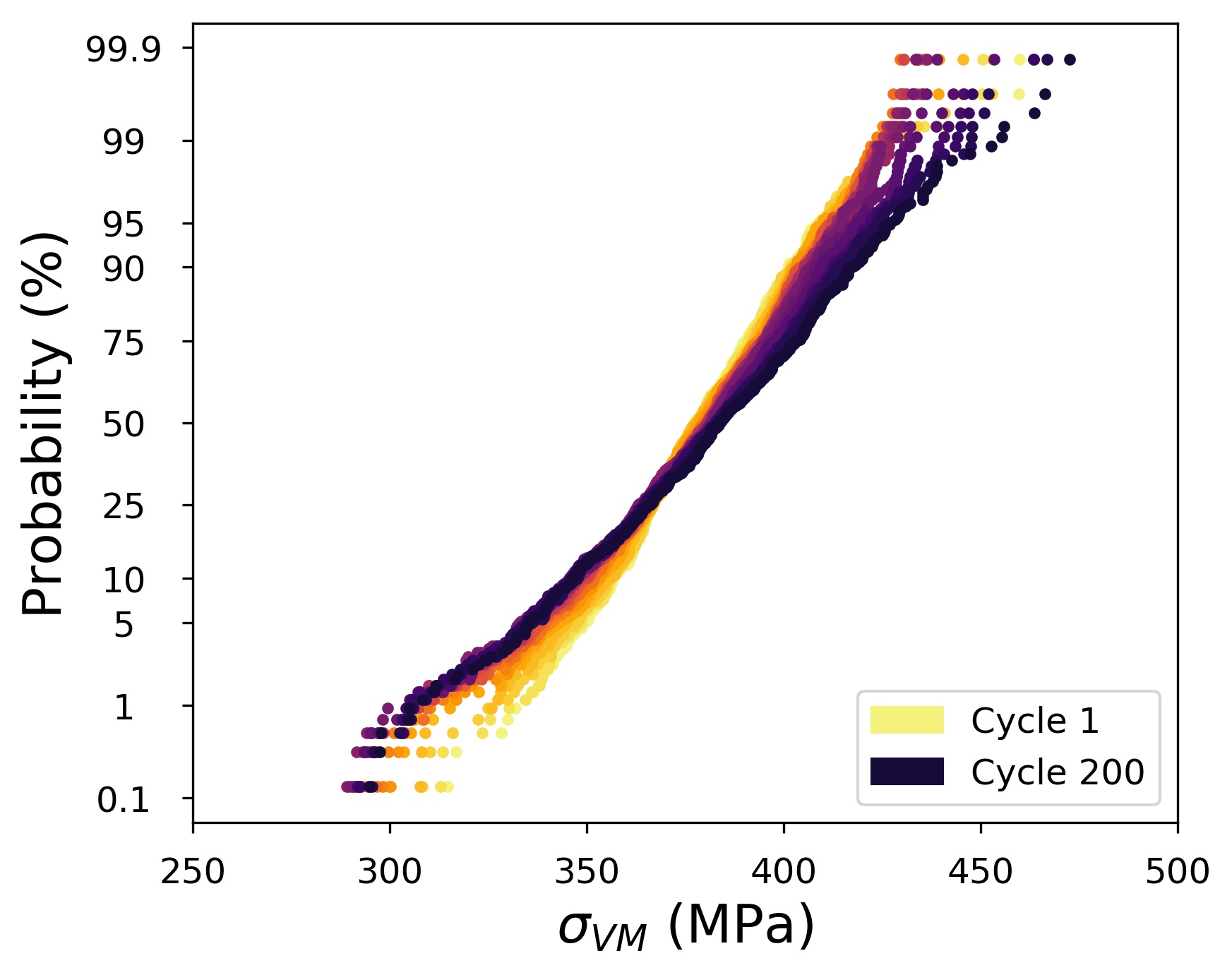}}
		\caption{The von Mises stress in the loaded state shows a rotation in the probability distribution around the 30\% value indicating that the distribution is spreading out and the median increases slightly but steadily with the accumulation of cycles. }
		\label{fig:VM_probplot}		
	\end{figure}

	Von Mises equivalent stress ($\sigma_{VM}$) is often used to determine when a ductile metal will yield. In the normal probability plot for the $\sigma_{VM}$ distribution shown in Figure~\ref{fig:VM_probplot}, the linear nature of the curve indicates a normal distribution. The rotation of the line through increasing cycles indicates that the distribution of $\sigma_{VM}$ is broadening as a function of cycle number (i.e. an increasing number of higher stress grains and lower stress grains). Additionally, the median value is increasing (slightly) through the cycles.
	
\begin{figure}[t]
		\centering
		\subfloat[]{\includegraphics[width=0.48\textwidth]{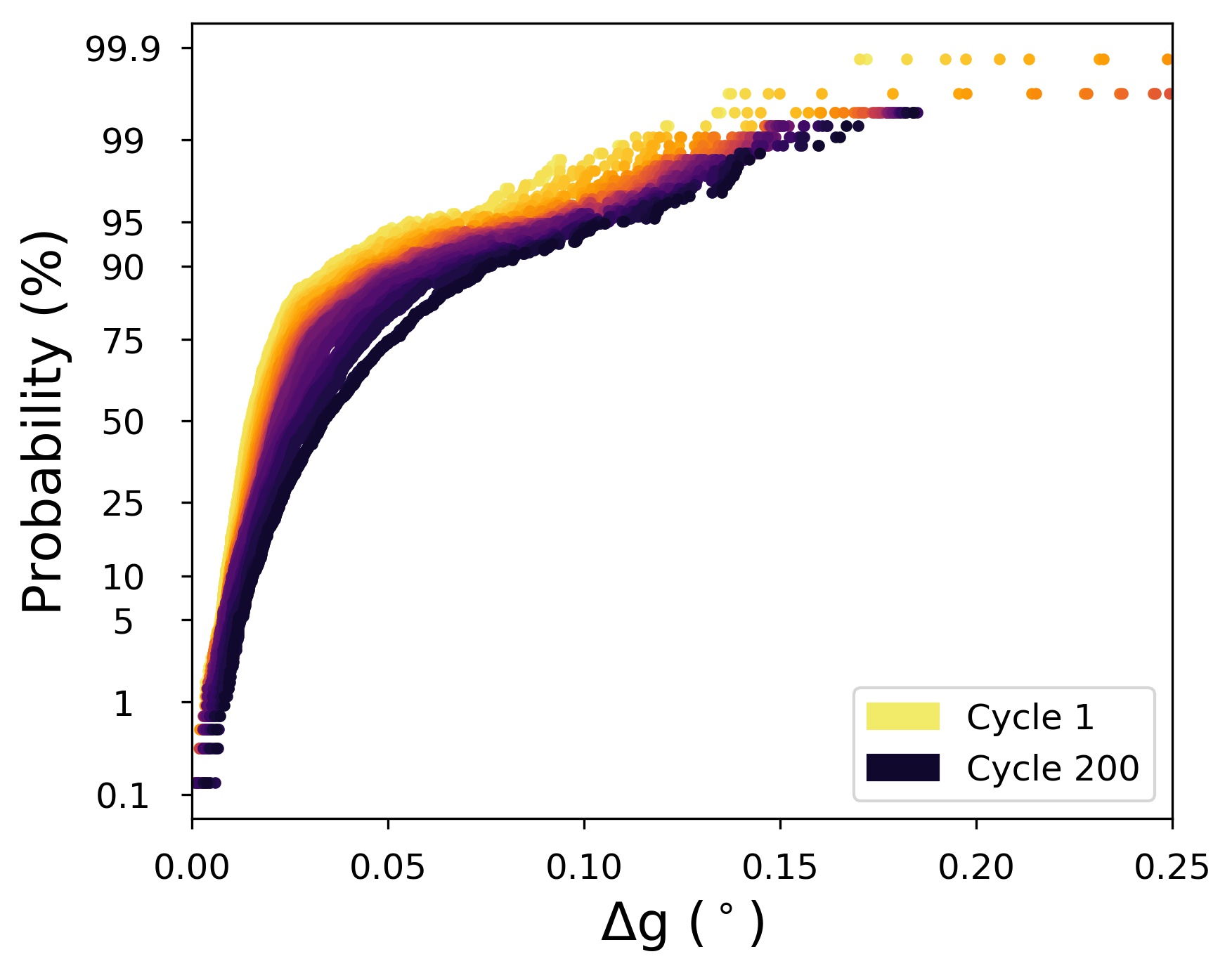}} \hspace{5pt}
		\subfloat[]{\includegraphics[width=0.48\textwidth]{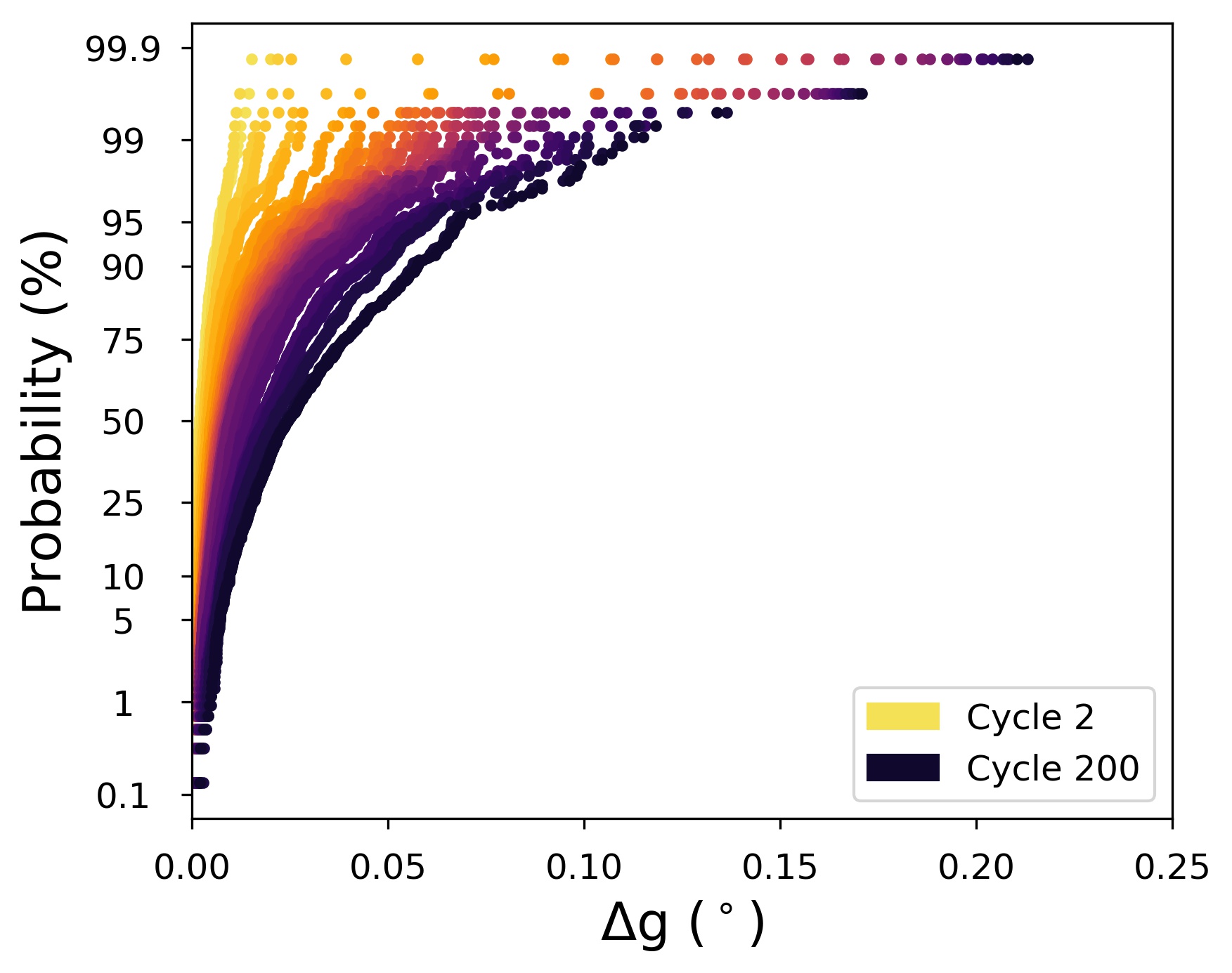}}
		\caption{(a) A probability distribution at each of the unloaded states was plotted for $\Delta g_0$, the change in orientation from the initial unloaded state. (b) A probability distribution was plotted for $\Delta g_1$, the change in orientation from the orientation after the first cycle to the other unloaded states in order to remove initial change from the yielding of the surface grains during the initial cycle. As the number of cycles increases, the distribution moves such that more grains have a higher $\Delta g$.}
		\label{fig:ori_probplot}		
	\end{figure}

	Grain rotations were calculated from the grain-averaged orientations to serve as a surrogate method for observing slip events. The normal probability plot in Figure~\ref{fig:ori_probplot} displays the change in grain orientation relative to initial grain orientation and shows that an increasing fraction of the grains have larger change in grain-averaged orientation ($\Delta g$) as cycles accumulate. Figure \ref{fig:ori_probplot}a is for $\Delta g_0$, the change in orientation from the initial unloaded state while Figure \ref{fig:ori_probplot}b is for $\Delta g_1$, the change in orientation from the unloaded state after the first cycle. Note that the uncertainty in change in orientation from scan to scan is estimated as approximately 0.003-0.005$\degree$, and is further addressed in Section~\ref{sec:discussion}.
	
\begin{figure}[h!]
		\centering
		\subfloat[]{\includegraphics[height=0.76\textheight]{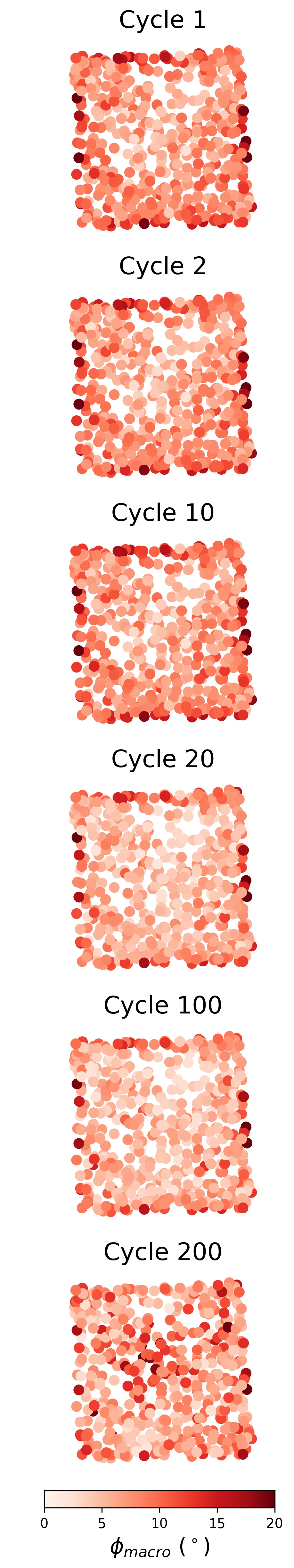}} \hspace{1pt}
		\subfloat[]{\includegraphics[height=0.76\textheight]{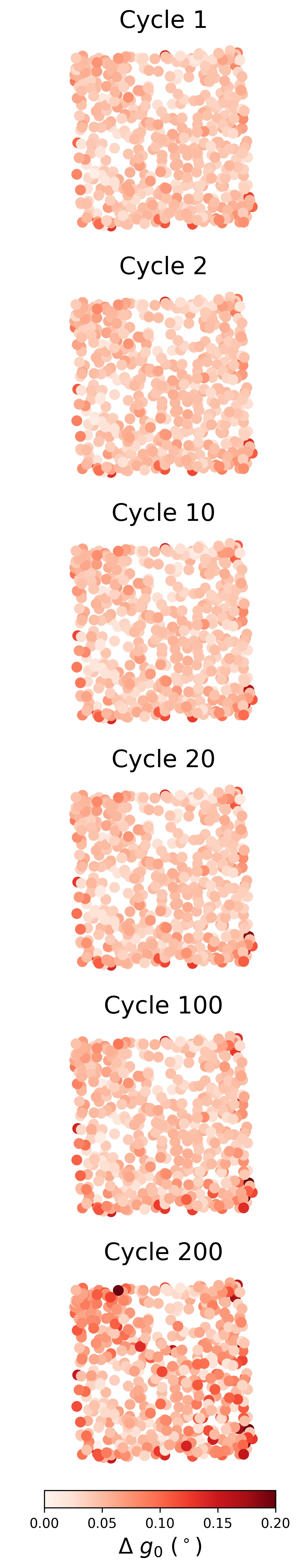}} \hspace{1pt}
		\subfloat[]{\includegraphics[height=0.76\textheight]{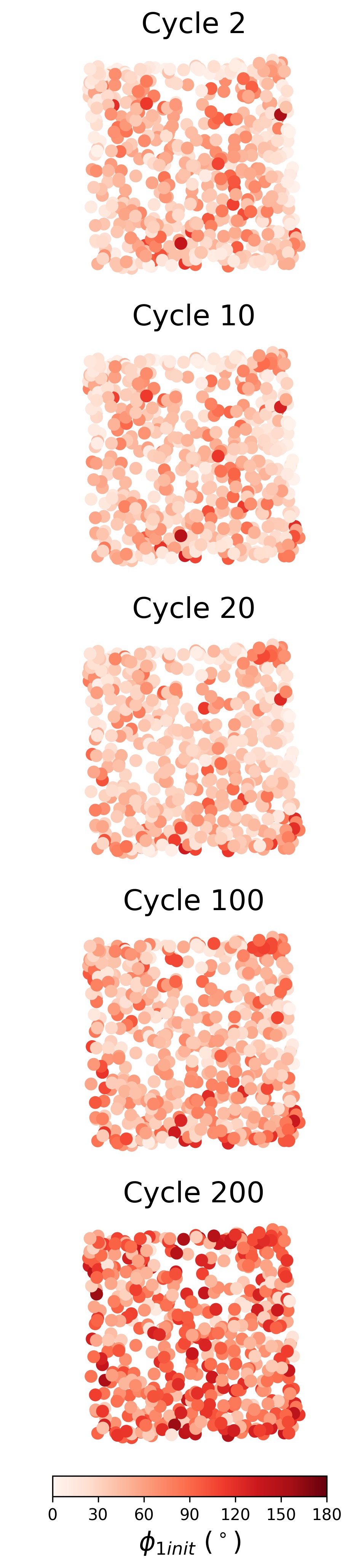}} \hspace{1pt}
		\subfloat[]{\includegraphics[height=0.76\textheight]{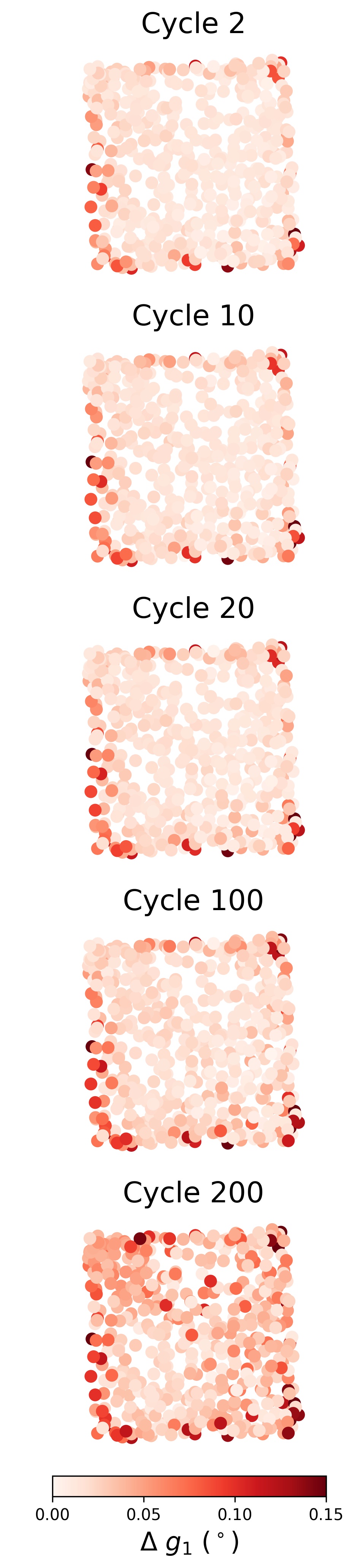}} \hspace{5pt}
		\subfloat{\includegraphics[width=0.09\textwidth]{xz.jpg}}
		\caption{In the loaded state, (a) $\phi_{macro}$ decreases with increasing cycles until cycle 100 with the surface grains having a higher $\phi_{macro}$. After cycle 100, the grains with high $\phi_{macro}$ are distributed throughout the sample. (b) As with the unloaded state, in the loaded state, the grain-averaged orientations show a larger $\Delta g$ from the initial loaded orientation during the $200^{th}$ cycle than during the 1st. In the unloaded state, (c) the coaxiality angle with respect to the stress tensor after the first cycle increases through the cycles. (d) The grain-averaged orientation is more misoriented from the initial orientation after the 200th cycle than after the 1st.}
		\label{fig:cycles_loaded2}		
	\end{figure}

	In addition to looking at $\sigma_{VM}$ and $\Delta g$, the stress tensor for each grain, $\bm{\sigma}_{grain}$, can be compared to the macroscopic stress tensor, $\bm{\sigma}_{macro}$. Stress coaxiality angle, $\phi_{macro}$ relates the stress state of an individual grain to the macroscopic applied load and can be calculated as:
	\begin{equation}
		\phi_{macro} = \cos^{-1} \left(\frac{\bm{\sigma}_{macro}:\bm{\sigma}_{grain}}{\|\bm{\sigma}_{macro}\|\|\bm{\sigma}_{grain}\|}\right).
	\end{equation}
	A low $\phi_{macro}$ indicates that a grain is well aligned with the the macroscopic stress tensor. In Figure \ref{fig:cycles_loaded2}a, $\phi_{macro}$ evolves with the number of cycles with the bulk grains having better alignment of their grain stress tensors with the macroscopic stress tensor than the surface grains until cycle 100. By cycle 200, the distribution of grains with high and low $\phi_{macro}$ are distributed throughout the sample showing no preference for the surface or bulk. Figure~\ref{fig:cycles_loaded2}b shows that $\Delta g_0$ from the initial orientation is negligible during the first 100 cycles, but by the $200^{th}$ cycle, there is a general increase in $\Delta g_0$ across the entire sample.
	
%	\begin{figure}[h!]
%		\centering
%		\subfloat[]{\includegraphics[height=0.7\textheight]{coaxiality_vertical_loaded.jpg}} \hspace{5pt}
%		\subfloat[]{\includegraphics[height=0.7\textheight]{misorientation_vertical_loaded.jpg}} \hspace{15pt}
%		\subfloat[]{\includegraphics[height=0.7\textheight]{coaxiality_vertical.jpg}} \hspace{5pt}
%		\subfloat[]{\includegraphics[height=0.7\textheight]{misorientation_vertical.jpg}} 
%		\subfloat{\includegraphics[height=0.02\textheight]{xz.jpg}}
%		\caption{In the loaded state, (a) $\phi_{macro}$ decreases with increasing cycles until cycle 100 with the surface grains having a higher $\phi_{macro}$. After cycle 100, the grains with high $\phi_{macro}$ are distributed throughout the sample. (b) As with the unloaded state, in the loaded state, the grain-averaged orientations show a larger $\Delta g$ from the initial loaded orientation during the $200^{th}$ cycle than during the 1st. In the unloaded state, (c) the coaxiality angle with respect to the stress tensor after the first cycle increases through the cycles. (d) The grain-averaged orientation is more misoriented from the initial orientation after the 200th cycle than after the 1st.}
%		\label{fig:cycles_loaded2}		
%	\end{figure}

	The coaxiality angle can also be calculated with respect to each grains' initial unloaded stress state, $\bm{\sigma}_{grain}^0$, rather than the macroscopic applied stress:
	\begin{equation}
	\phi_{init} = \cos^{-1} \left( \frac{\bm{\sigma}_{grain}^0:\bm{\sigma}_{grain}}{\|\bm{\sigma}_{grain}^0\|\|\bm{\sigma}_{grain}\|}\right).
	\end{equation}
	In this case, $\phi_{init}$ is a measure of the how well aligned a grain's stress tensor is with its stress tensor in the initial state. Due to the major changes in stress and orientation that occurred during the first cycle, the subsequent unloaded states were compared to the sample state after the first cycle. As the number of cycles increases in Figure \ref{fig:cycles_loaded2}c, the coaxiality angle increases, indicating that the stress tensor of each grain is becoming less and less aligned with its initial stress tensor. In Figure \ref{fig:cycles_loaded2}d, the 2nd cycle has little change in orientation from after the first cycle, but, again, by the $200^{th}$ cycle, $\Delta g_1$ has generally increased across the grain ensemble.

%	\begin{figure}[h!]
%		\centering
%		\subfloat[]{\includegraphics[height=0.825\textheight]{coaxiality_vertical.jpg}} \hspace{5pt}
%		\subfloat[]{\includegraphics[height=0.825\textheight]{misorientation_vertical.jpg}} \hspace{5pt}
%		\subfloat{\includegraphics[height=0.15\textheight]{xz.jpg}}
%		\caption{In the unloaded state, (a) the coaxiality angle with respect to the stress tensor after the first cycle increases through the cycles. (b) The grain-averaged orientation is more misoriented from the initial orientation after the 200th cycle than after the 1st.}
%		\label{fig:cycles_loaded2}b		
%	\end{figure}

\subsection{Grain Rotations}
\begin{figure}[b!]
		\centering
		\subfloat[]{\includegraphics[width=0.32\textwidth]{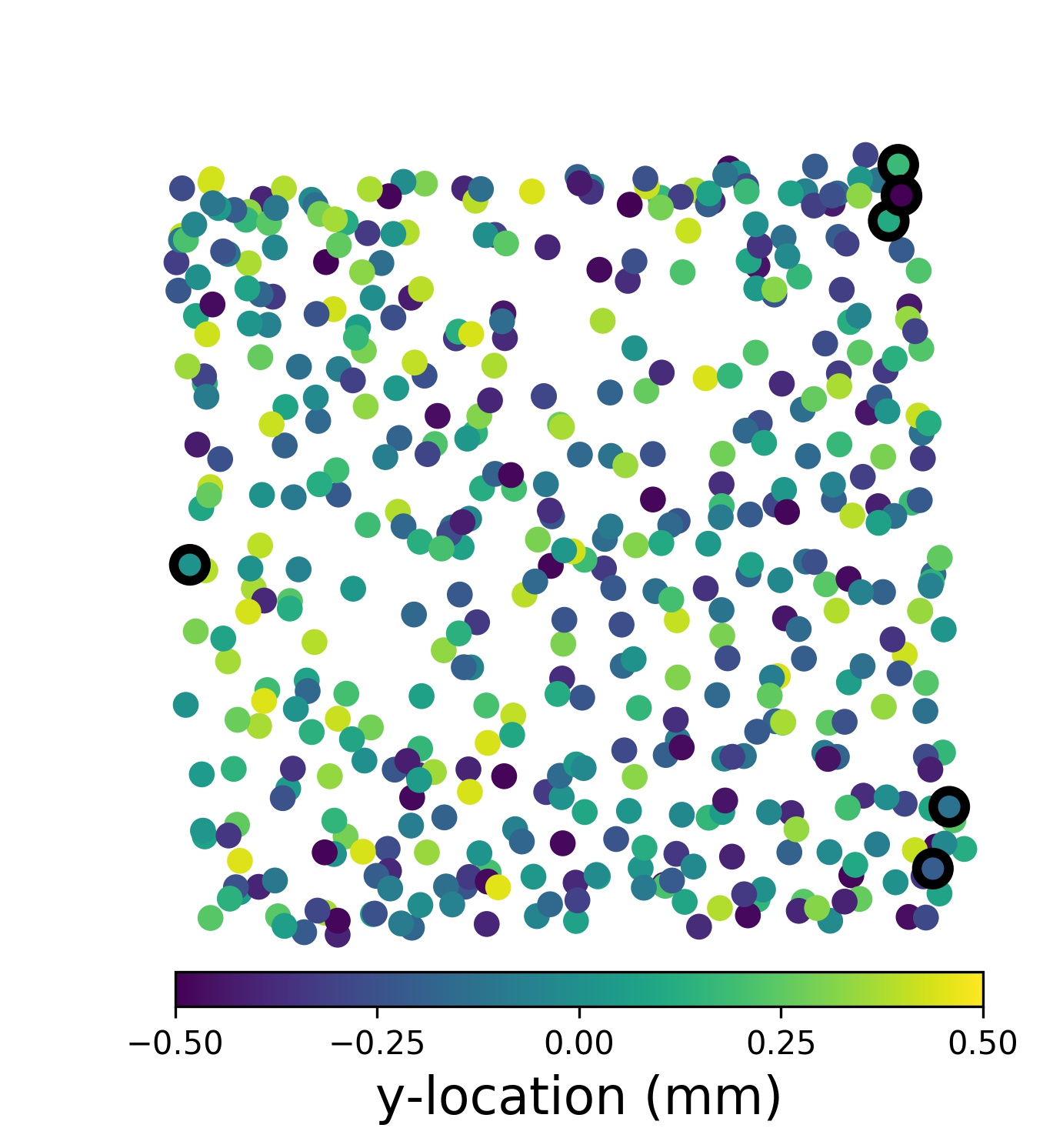}} \hspace{5pt}
		\includegraphics[height=0.1\textheight]{xz.jpg}\\
		\subfloat[Grain 166]{\includegraphics[width=0.25\textwidth]{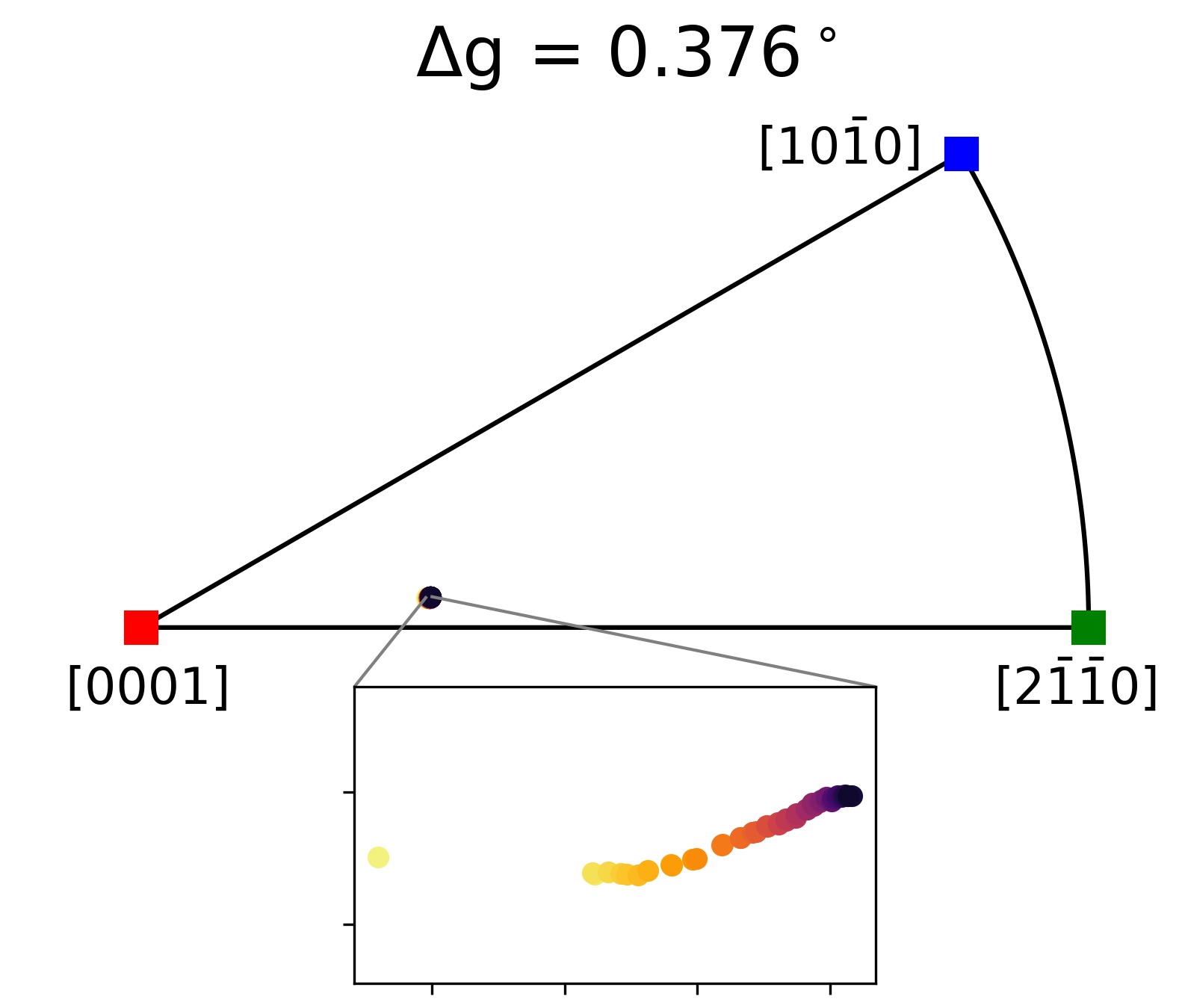}} \hspace{5pt} 
		\subfloat[Grain 167]{\includegraphics[width=0.25\textwidth]{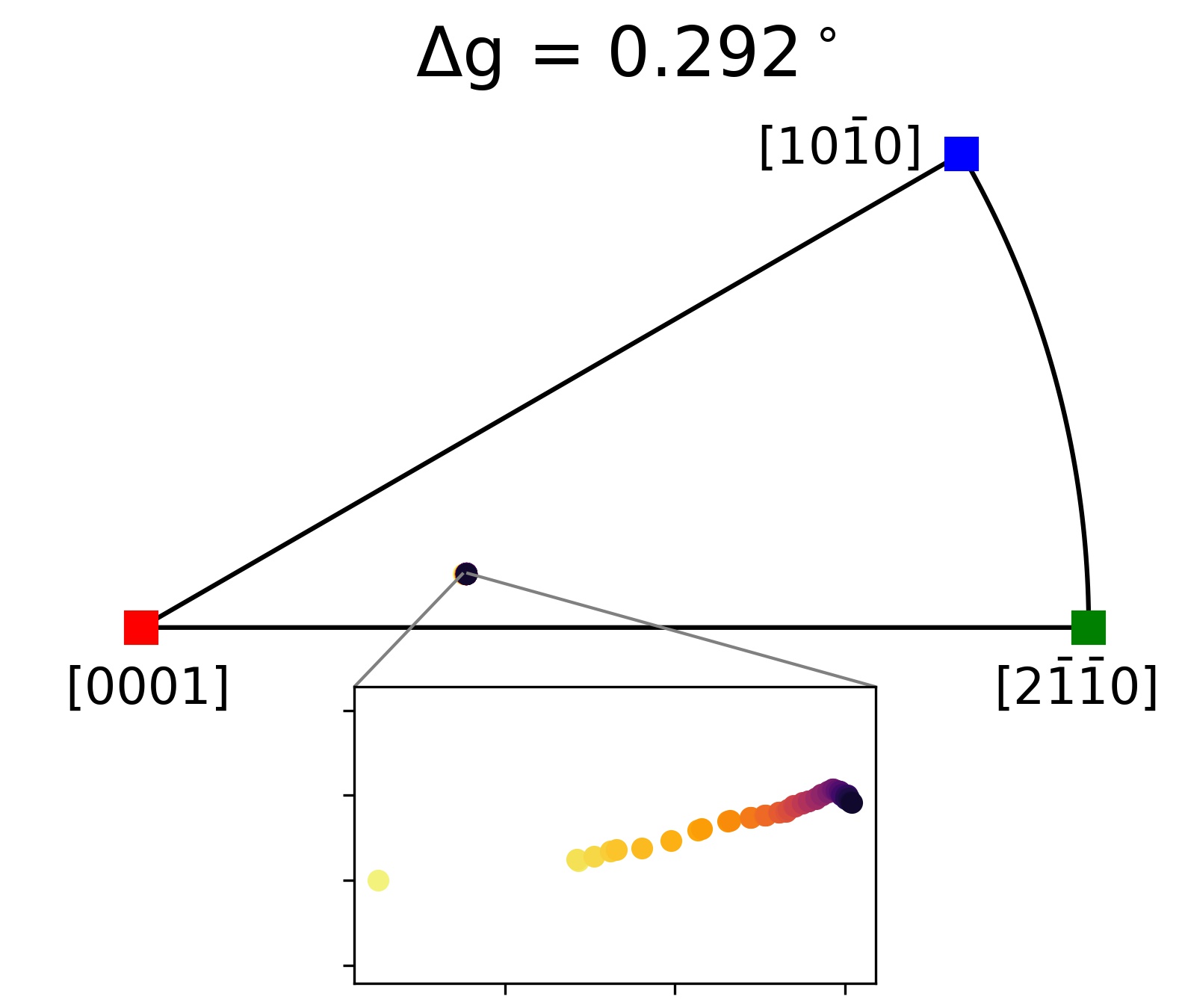}} \hspace{5pt}
		\subfloat[Grain 347]{\includegraphics[width=0.25\textwidth]{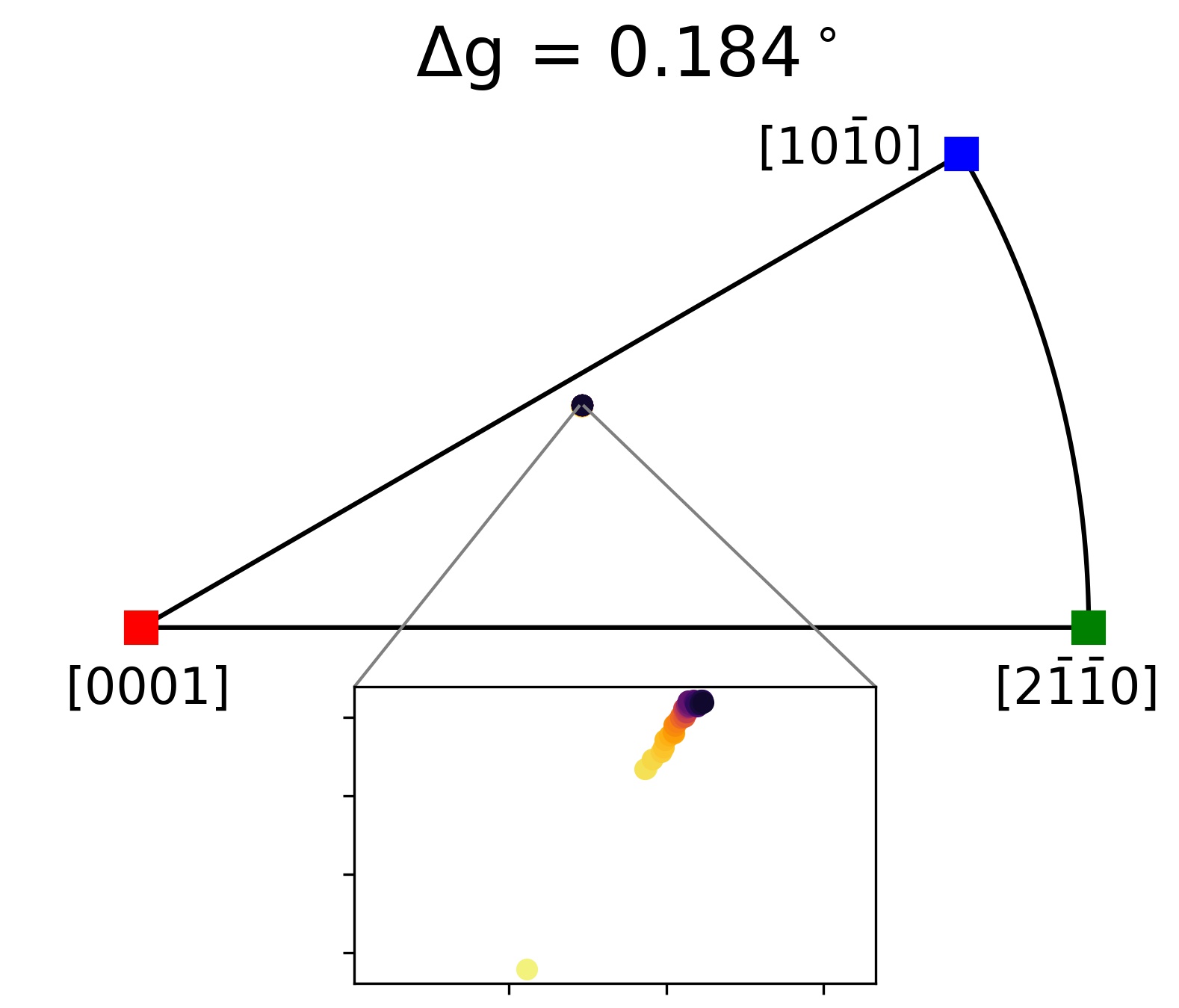}} \\
		\subfloat[Grain 62]{\includegraphics[width=0.25\textwidth]{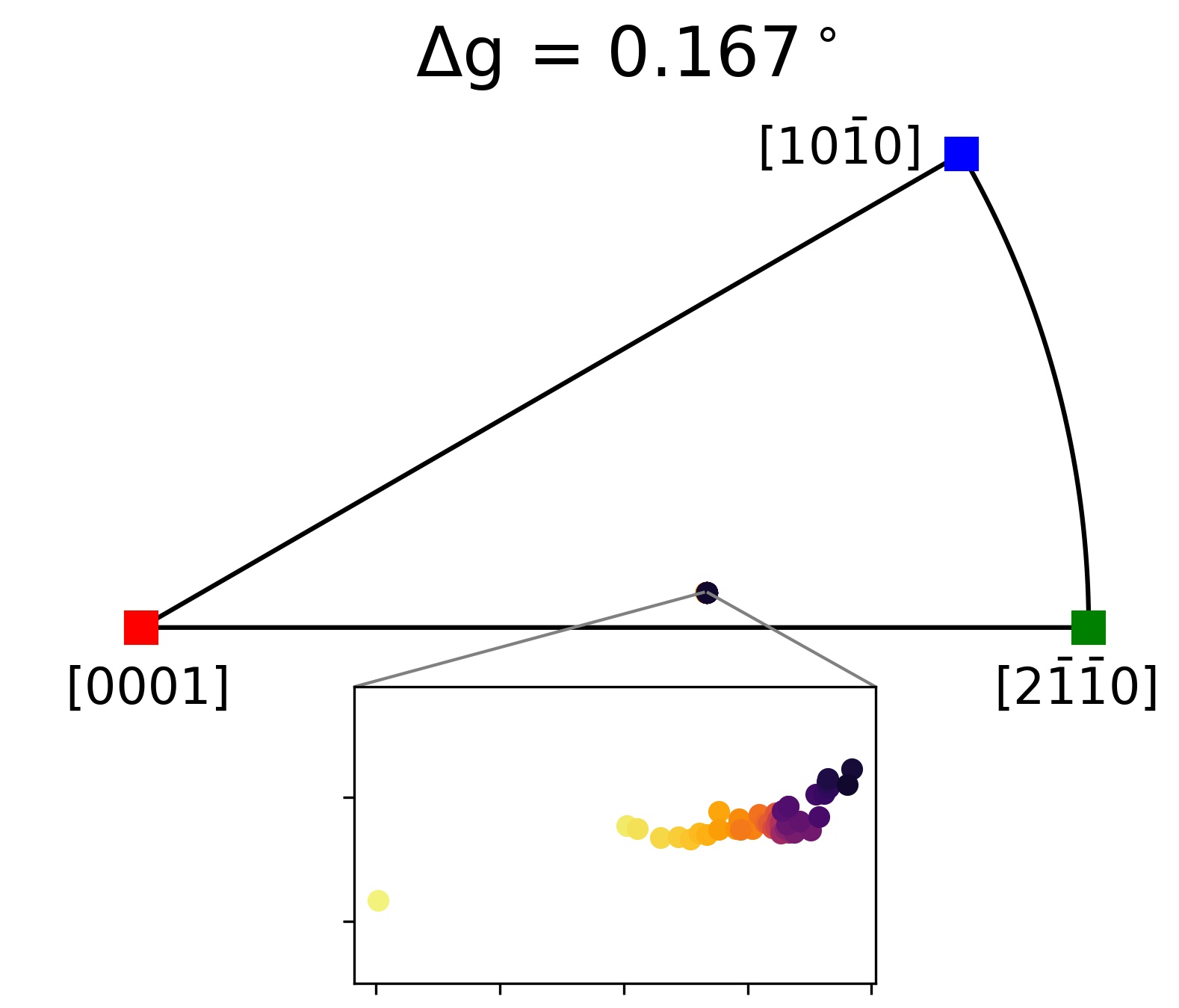}} \hspace{5pt}
		\subfloat[Grain 66]{\includegraphics[width=0.25\textwidth]{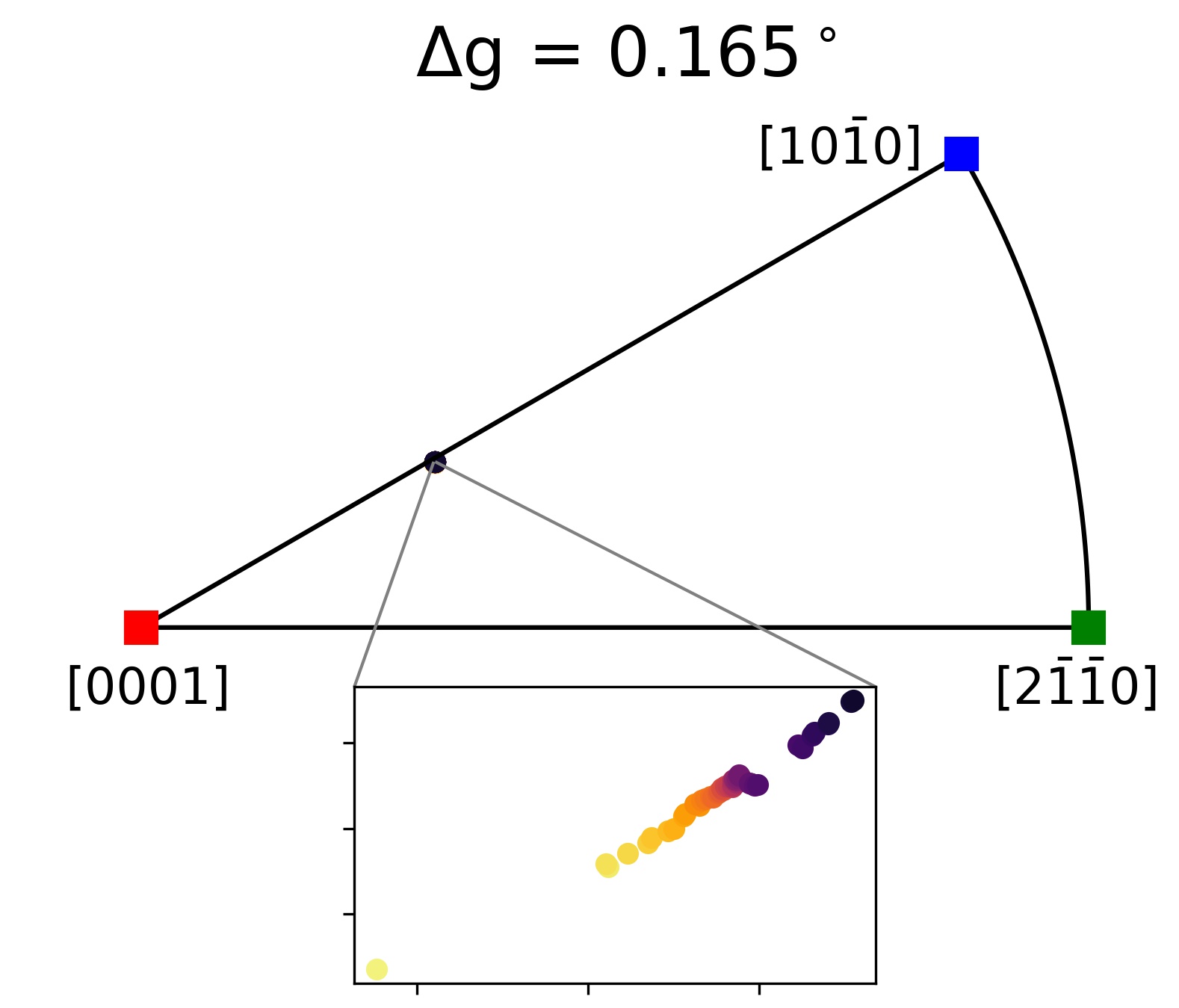}}\hspace{5pt}
		\subfloat[Grain 148]{\includegraphics[width=0.25\textwidth]{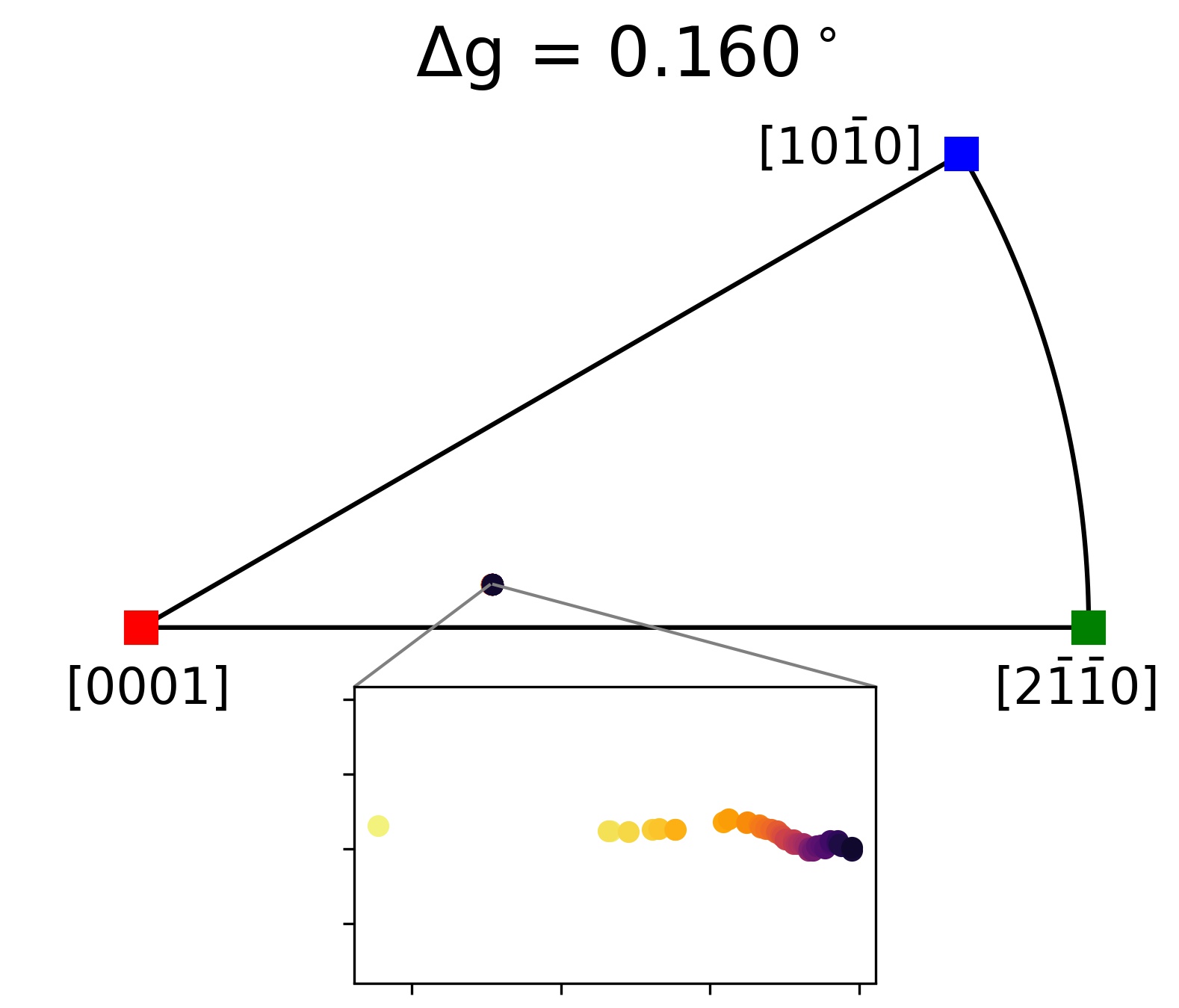}}
		
		\caption{(a) Location of the grain COMs in the x-z plane colored according to their location in the y-dimension marked with the six grain with the largest change in grain-averaged orientation ($\Delta g_0$). (b-g) The six grains rotate so that their c-axis points further away from the loading direction. $\Delta g$ is greater during the first cycle than any of the subsequent ones.}
		\label{fig:ipfs}		
	\end{figure}

\begin{figure}
		\includegraphics[width=0.8\textwidth]{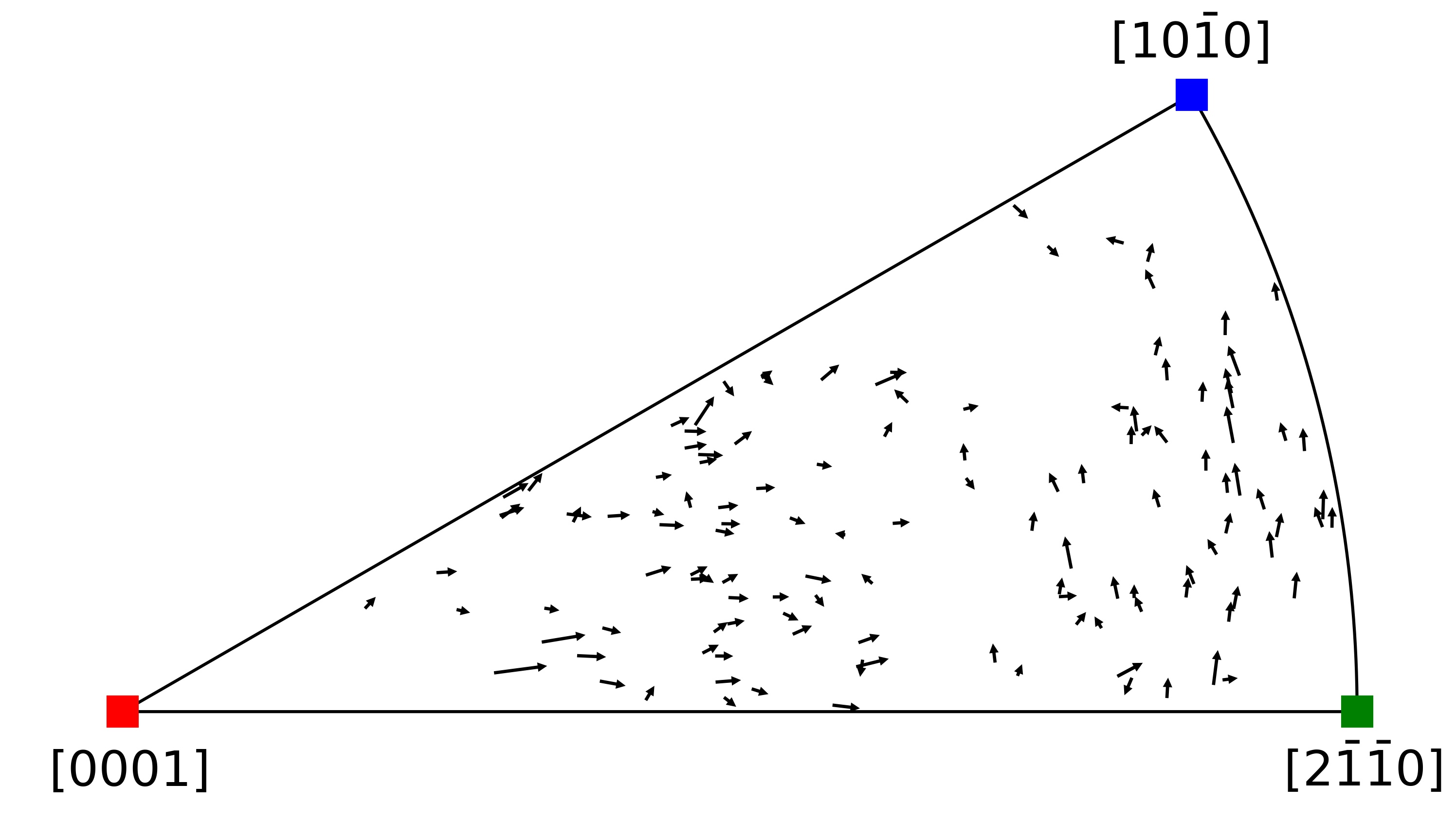}
		\caption{IPF triangle for the loading axis plotted with arrows pointing in the direction of the orientation change (at 10$\times$ the magnitude of change) for grains with $> 0.05 \degree$ of orientation change from the initial state to the final state.}
		\label{fig:ipf_arrows}
	\end{figure}

	 Investigating the outlier grains further, the six grains with $\Delta g_0 > 0.15\degree$ are marked in Figure \ref{fig:ipfs}a, and are clearly all located on the surface of the sample. When the orientations for these grains are plotted for each of the cycles in Figure \ref{fig:ipfs}b-g, it becomes evident that all six of of the grains are rotating such that their c-axis points away from the loading direction. In following that line of thought, the grains with $\Delta g_0 >0.05 \degree$ are plotted on an IPF triangle in Figure \ref{fig:ipf_arrows} with the direction of the arrow indicating the direction of the grain rotation and the length of the arrow equaling 10x the magnitude of $\Delta g_0$.

\section{Discussion} \label{sec:discussion}
\subsection{Aggregate Evolution}

	The tensile strains present in the initial state seen in Figure \ref{fig:strain22_probplot}c can be attributed to the recast layer from the EDM \cite{SrinivasaRao2016}. Not only does this speak to the precision of the measurements in this dataset, but also causes some interesting behavior for the first cycle. The residual strain present along the surface of the sample in the initial state goes away after the first cycle indicating that the grains with high residual $\epsilon_{yy}$ yielded during the first cycle, thereby relieving their residual strain (and stress).
	
	Orientation precision for the values in Figures \ref{fig:strain22_probplot}d, \ref{fig:ori_probplot}, \ref{fig:cycles_loaded2}b, \ref{fig:ipfs}, and \ref{fig:ipf_arrows} can be considered to be $\sim$0.003$\degree$ based on Bernier et al. \cite{Bernier2011}. However, this value is based on a series of ruby scans done for only a 120$\degree$ sweep rather than the 360$\degree$ sweep done here, thus the uncertainty in orientation is likely to be even smaller than $\sim$0.003$\degree$ for the full rotation. Work is being pursued on a new set of ruby experiments with a variety of experimental parameters, including a full 360$\degree$ sweep, in order to refine the determination of the precision of f{}f-HEDM experimental results with current instrumentation and setup.
	
	As grains slip, they rotate to accommodate the slight changes in shape and stress state, and lattice rotation will occur subsequently to maintain compatibility, thus, grain rotation can be used as a proxy for slip. Consequently, as grains continue to slip through subsequent cycles, the grain rotation serves as an indicator of the accumulation of plastic strain and possibly damage in a grain. This is significant as the sample was being cycled at only $90~\%$ of its yield stress (Fig. \ref{fig:ti7_stress_strain}) and consistently exhibited macroscopic elastic behavior with negligible hysteresis. Therefore, although the sample appeared to deform only elastically from a macroscopic perspective, changes in stress state and orientation were occurring heterogeneously in the individual grains. In the absence of nucleation at microstructural heterogeneities such as voids and brittle particles, single phase metals typically form fatigue cracks at slip steps or grain boundaries. The significance of the continual slip is therefore that it leads to the build-up of damage that drives the initiation and growth of fatigue cracks and eventually failure.  Another way to state this is that irreversible dislocation motion occurred on a microscopic scale in a majority of grains.  
	
\begin{figure}[t!]
	 	\centering
	 	\includegraphics[width=0.48\textwidth]{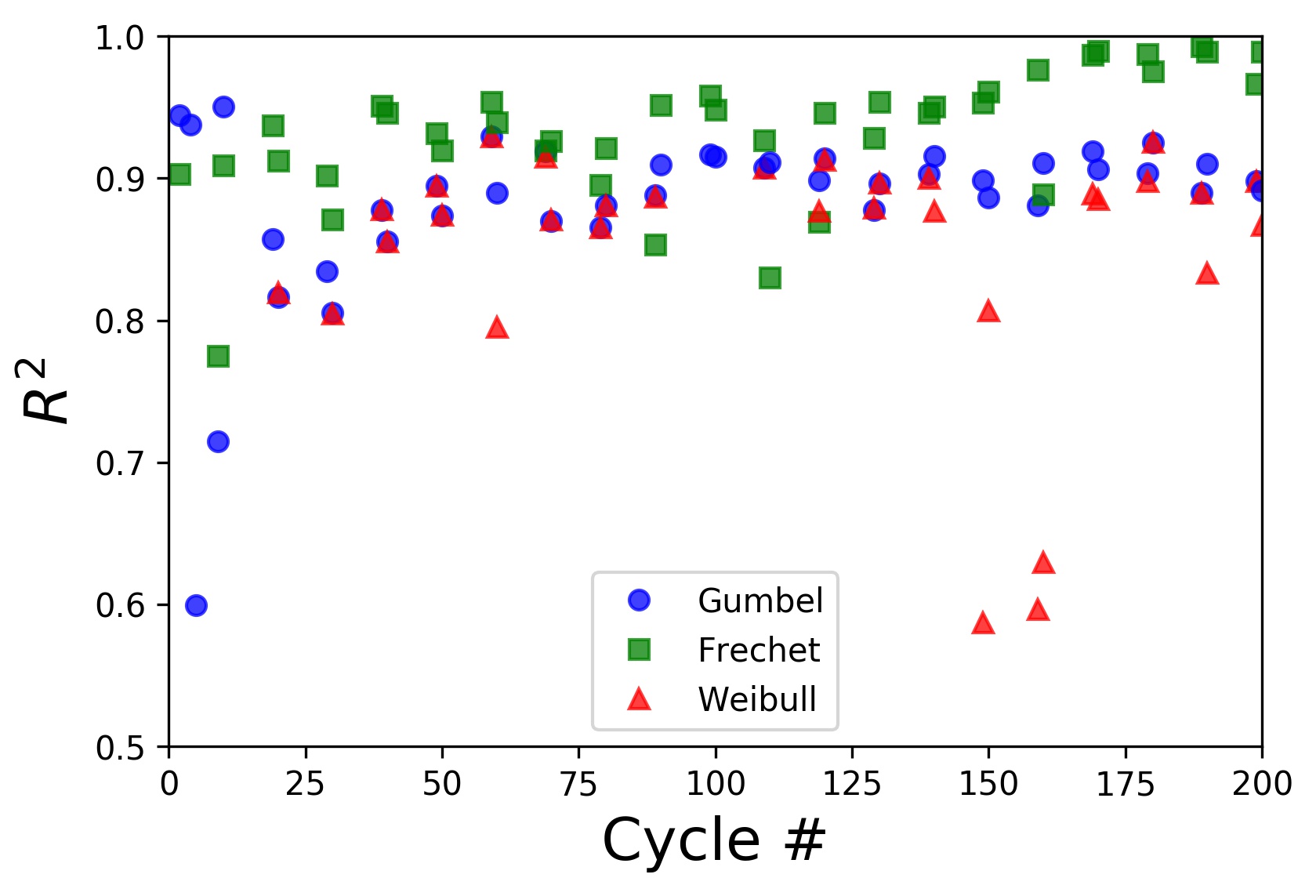}
	 	\caption{The three extreme value distributions were fit to the upper tail of the $\Delta g$ probability distributions, and the Fr\'{e}chet distribution consistently fits the best.}
	 	\label{fig:extreme_values}
	 \end{figure}

	 Accordingly, the upper tails of the distributions of $\Delta g$, defined here as having $> 0.01\degree$ change in orientation when compared to the orientation after the first cycle, were analyzed using extreme value statistics. The Fisher-Tippett-Gnedenko theorem \cite{Fisher1928,Gnedenko1943} states that the probability distribution function for a convergent generalized extreme value distribution will fit one of three distributions: Gumbel \cite{Gumbel1935,Gumbel2012}, Fr\'{e}chet \cite{Frechet1927}, or Weibull \cite{Weibull1951}. These three distributions were fit to the upper tails of $\Delta g$, and Figure \ref{fig:extreme_values} shows the $R^2$ values for the fits at dif{}ferent cycle numbers and reveals that the Fr\'{e}chet distribution (Eq. \ref{eq:frechet}) consistently fits the tail better than the other two. The Fr\'{e}chet probability distribution function has form:
	 \begin{equation} \label{eq:frechet}
	 	f(x) = \frac{\alpha}{\beta} \left(\frac{\beta}{x-\gamma}\right)^{\alpha-1} \exp\left(-\left(\frac{\beta}{x-\gamma}\right)^\alpha\right)
	 \end{equation} 
	 where $\alpha>0$ is the shape parameter, $\beta>0$ is the scale parameter, and $\gamma$ is the location parameter. The fit for the last 50 cycles was averaged and found to be $\alpha = 0.952$, $\beta = 0.018$, and $\gamma = 0.010$. This analysis clarifies that the distribution of plastic strain, as detected by the orientation change was far from Gaussian and exhibited a heavy upper tail.  This indicates that a minority of grains experienced much larger plastic behavior than the rest, which helps us to understand how fatigue crack initiation is an extreme event \cite{Bazant2017,Castillo2004}.  To the authors' knowledge, this is the first time that the heterogeneity of slip accumulation under nominally elastic cyclic loading has been quantified to this level of detail.
	 
\subsection{Grain Rotations and Slip}
	
	The six grains with the largest changes in orientation are oriented such that the basal $\langle$a$\rangle$ and/or pyramidal $\langle$c+a$\rangle$ systems are activating and rotate in the direction that increases the angle between their c-axis and the loading direction (Fig. \ref{fig:ipfs}). This rotation orients the grains in such a way as to promote prismatic $\langle a \rangle$ slip, the softest of the systems available for deformation in Ti-7Al. As a whole, the grains in the sample are rotating in such a way that a tension texture begins to develop \cite{Wenk2000,Kouchmeshky2009, Huang2011,Lind2013,Chapuis2015}, albeit the process is in its early stages. Throughout the sample, the grains near [0001] are slipping on basal $\langle$a$\rangle$ and pyramidal $\langle$c+a$\rangle$ systems and rotating so that the [0001] axis points away from the loading direction. Additionally, the grains far from [0001] are slipping on the prismatic $\langle$a$\rangle$ system and rotating so that the $[10\bar{1}0]$ direction becomes more aligned to the loading direction.
	
\begin{figure} [t]
		\centering
		\subfloat[]{\includegraphics[width=0.4\textwidth]{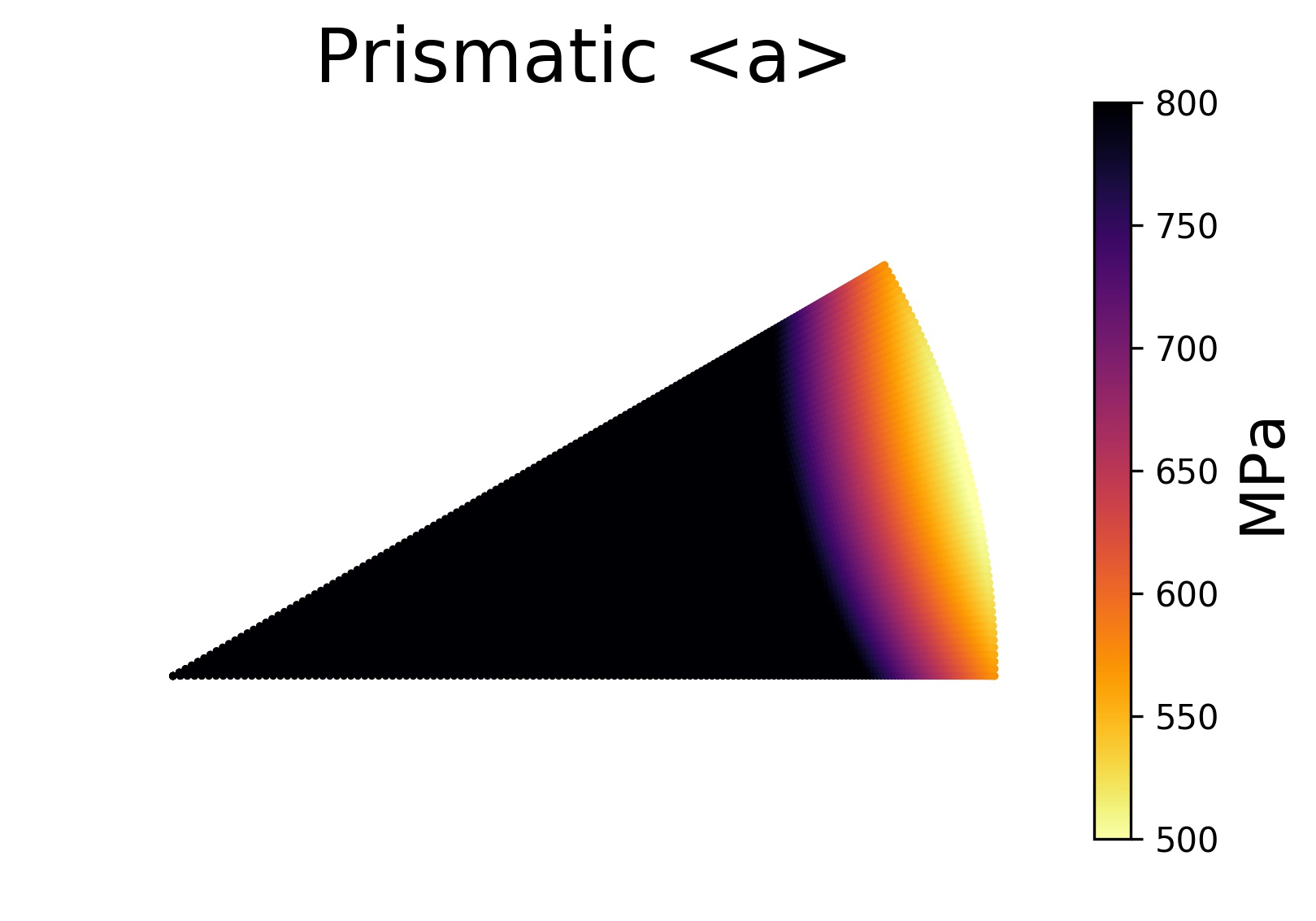}} \hspace{5pt}
		\subfloat[]{\includegraphics[width=0.4\textwidth]{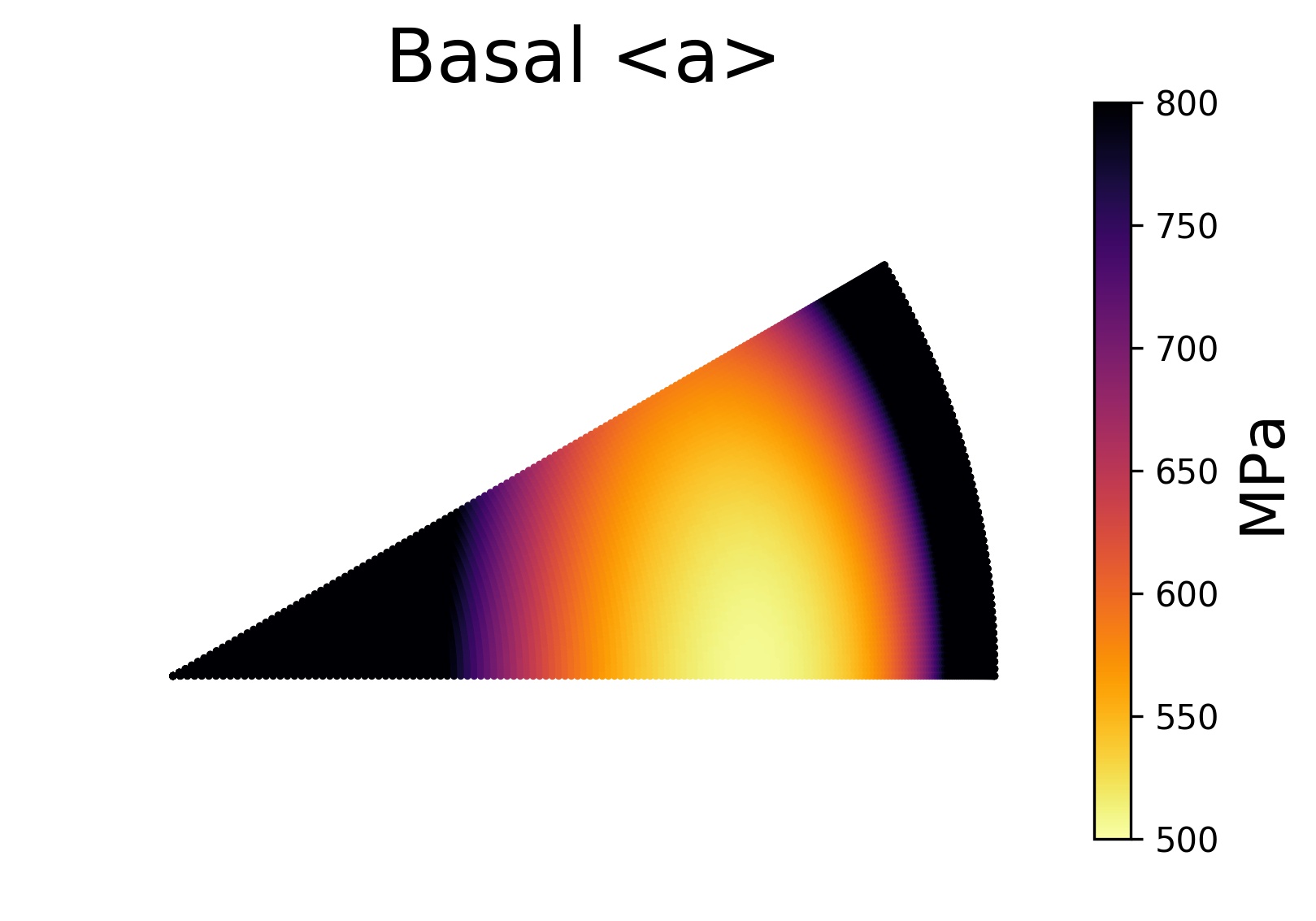}} \\
		\subfloat[]{\includegraphics[width=0.4\textwidth]{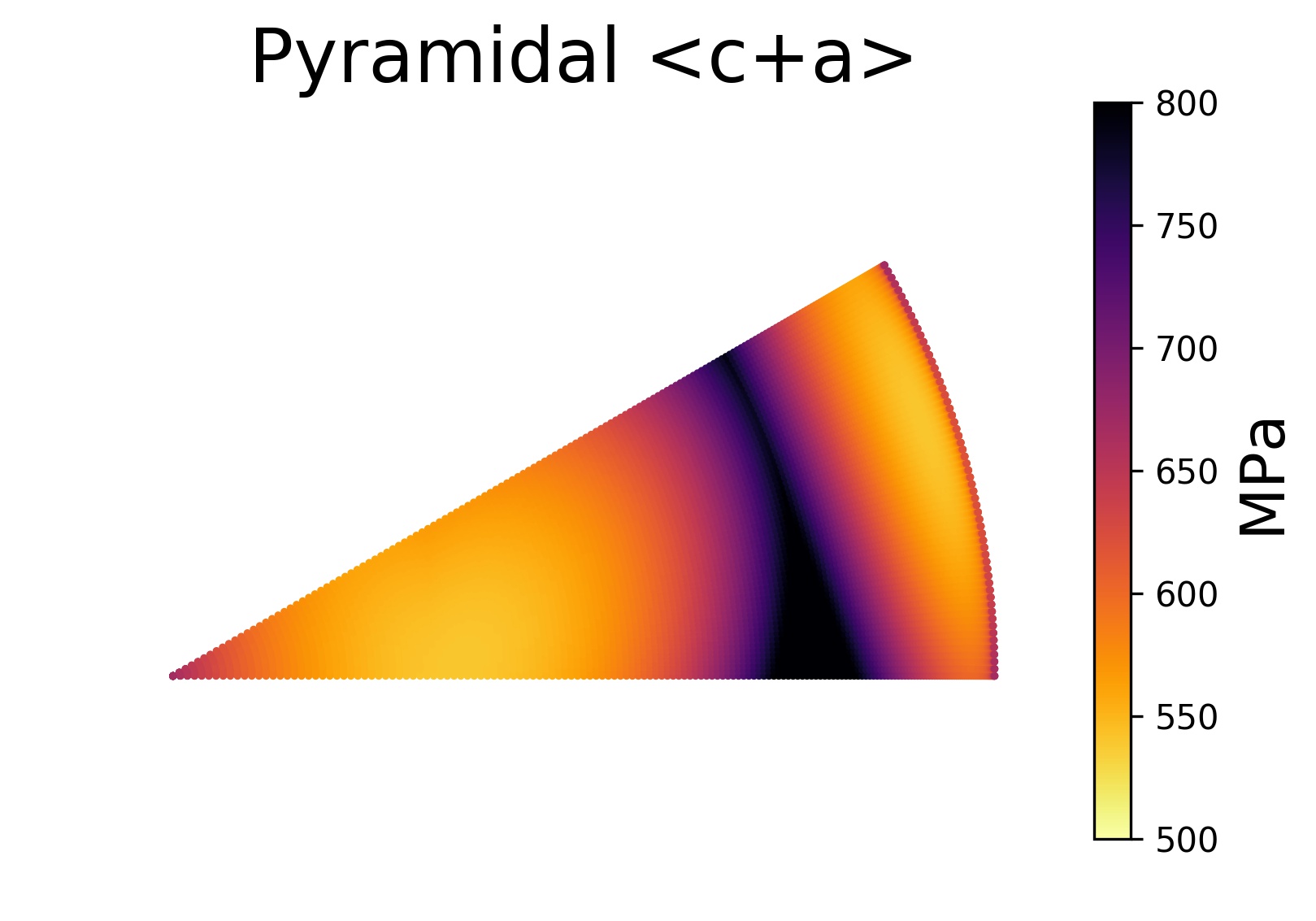}}
		\caption{(a-c) The CRSS for each slip system was divided by the maximum Schmid factor at a given orientation on the fundamental triangle to give the uniaxial stress required to activate the slip system at that orientation.}
		\label{fig:IPF_uniaxial}
	\end{figure}

	The CRSS (see Table \ref{tab:slip systems}) for each slip system was divided by the Schmid factor at a given orientation to show the uniaxial stress required to activate each of the slip systems relative to a grain's orientation. It is also important to note, however, that no grain will be in a purely uniaxial stress state, but this analysis gives an idea of what is generally occurring. As can be seen in Figure \ref{fig:IPF_uniaxial}, the grains oriented on the right side of the triangle with [0001] far from the loading direction require a lower stress to activate the prismatic $\langle$a$\rangle$ slip systems while the grains with orientations in the middle of the triangle are more likely to slip on the basal $\langle$a$\rangle$ family. Moreover, the pyramidal $\langle$c+a$\rangle$ systems must activate for the grains oriented with their c-axis parallel to the loading direction. The pyramidal $\langle$a$\rangle$ is not shown since it is rarely observed \cite{Bridier2008}.

\begin{figure}[t]
	\centering
	\subfloat[]{\includegraphics[width=0.65\textwidth]{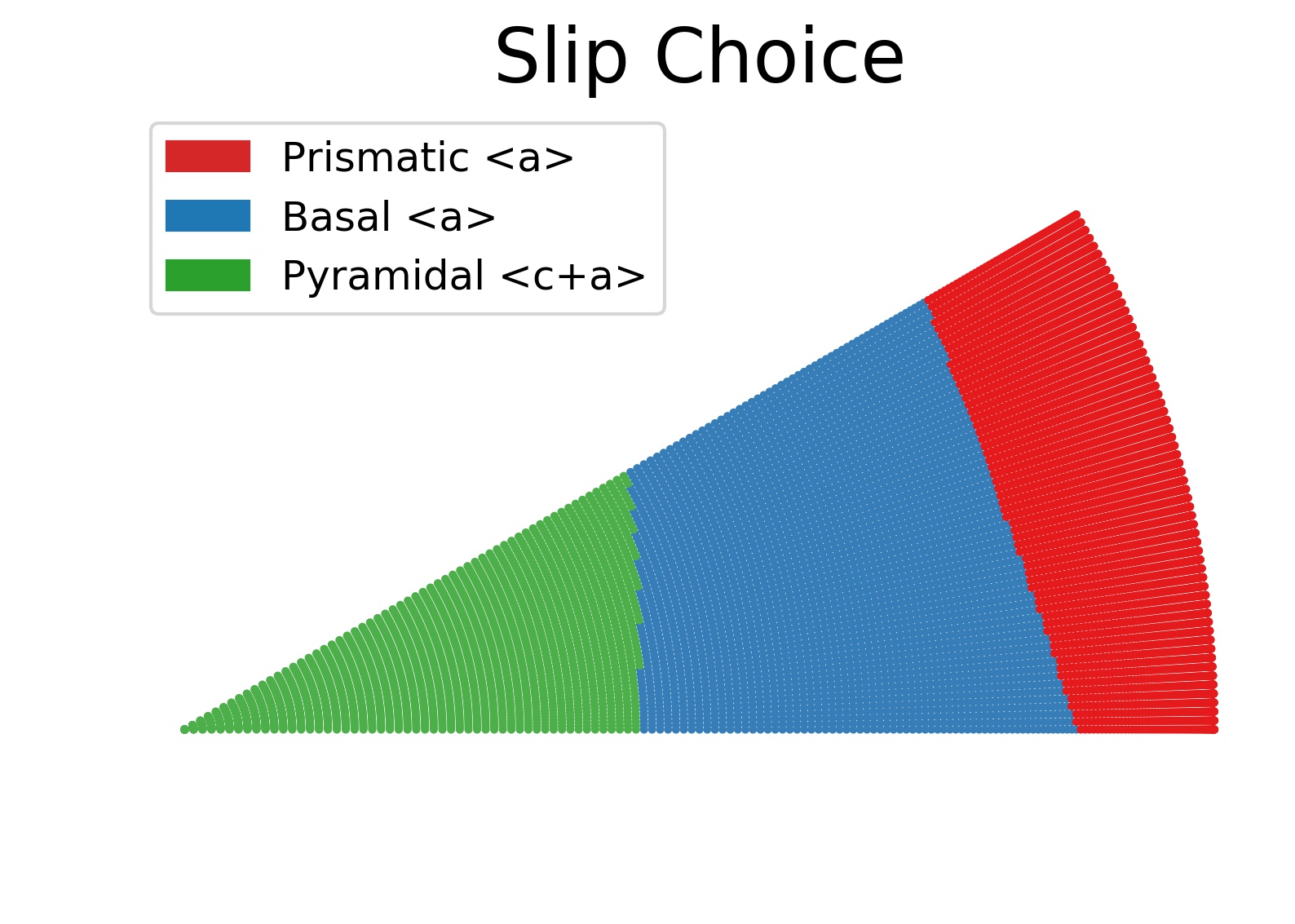}}
	\caption{Each region of the fundamental triangle is colored according to the preferred slip system at that orientation based on Schmid factor (Fig. \ref{fig:IPF_uniaxial}) and the CRSS calculated in Pagan et al. \cite{Pagan2017}. The regions for the preferred slip systems were adjusted to reflect slip choice based on understanding of hardening/softening behavior and slip rate sensitivity dependence \cite{Williams1968,Beausir2007,May2010}.}
	\label{fig:slip_choice}
	\end{figure}

	The uniaxial stresses in Figure \ref{fig:IPF_uniaxial} are taken and combined into one triangle to better show the likely slip system to be activated at a given orientation. However, the slip rate dependence and slip system strength evolution are not taken into account by the CRSS which assumes rate independent plasticity. In terms of rate sensitivity, Beausir et al. \cite{Beausir2007} calculated that the the prismatic $\langle$a$\rangle$ and basal $\langle$a$\rangle$ systems have a higher rate sensitivity, which functionally means that they slip more easily at low stresses, than the pyramidal $\langle$c+a$\rangle$ slip systems. Additionally, Pagan et al. \cite{Pagan2017} found that the pyramidal $\langle$c+a$\rangle$ slip systems harden rapidly while the prismatic $\langle$a$\rangle$ and basal $\langle$a$\rangle$ systems soften. It was also noted that the CRSS values for the basal $\langle$a$\rangle$ and pyramidal $\langle$c+a$\rangle$ slip systems reported in Pagan et al. are lower than those reported for titanium alloys with similar aluminum content (Ti-6.7Al) \cite{Williams1968,May2010}. The slip choice triangle was adjusted based on this rationale and recalculated with higher CRSS values (Table \ref{tab:slip systems_CRSS}) for the basal $\langle$a$\rangle$ and pyramidal $\langle$c+a$\rangle$ slip systems yielding the result in Figure \ref{fig:slip_choice}. In light of these adjustments, Figure \ref{fig:ipf_arrows} can be reexamined with respect to the slip system preferences in Figure \ref{fig:slip_choice}. As stated previously the grains with orientations on the right side of the triangle which are indeed slipping on the prismatic slip systems and rotating around the c-axis while the grains with orientations in the middle of the triangle are rotating so that their c-axis points away from the loading direction.
	
\begin{table}[t!]
	\centering
	\begin{tabular}{c|c|c}
		\textbf{Slip System}  & \textbf{CRSS from Pagan et al.}  & \textbf{Adjusted CRSS} \\ \cline{1-3}
		Prismatic $\langle$a$\rangle$   & 248 MPa & 248 MPa \\
		Basal $\langle$a$\rangle$  & 253 MPa & 293 MPa *\\
		1st Order Pyramidal $\langle$c+a$\rangle$    & 270 MPa & 365 MPa **\\
	\end{tabular}\par
	\justifying
	\vspace{8pt}
	\noindent\** from May \cite{May2010}. \\
	\noindent\*** average of May \cite{May2010} and Pagan et al. \cite{Pagan2017}.
	\caption{The CRSS as reported in Pagan et al. and the CRSS values adjusted for hardening behavior and rate dependence used in Figure \ref{fig:slip_choice}.}
	\label{tab:slip systems_CRSS}
\end{table}

\subsection{Ef{}fect of Orientation on Micromechanics}
\begin{figure}[h]
		\centering
		\subfloat{\includegraphics[width=0.48\textwidth]{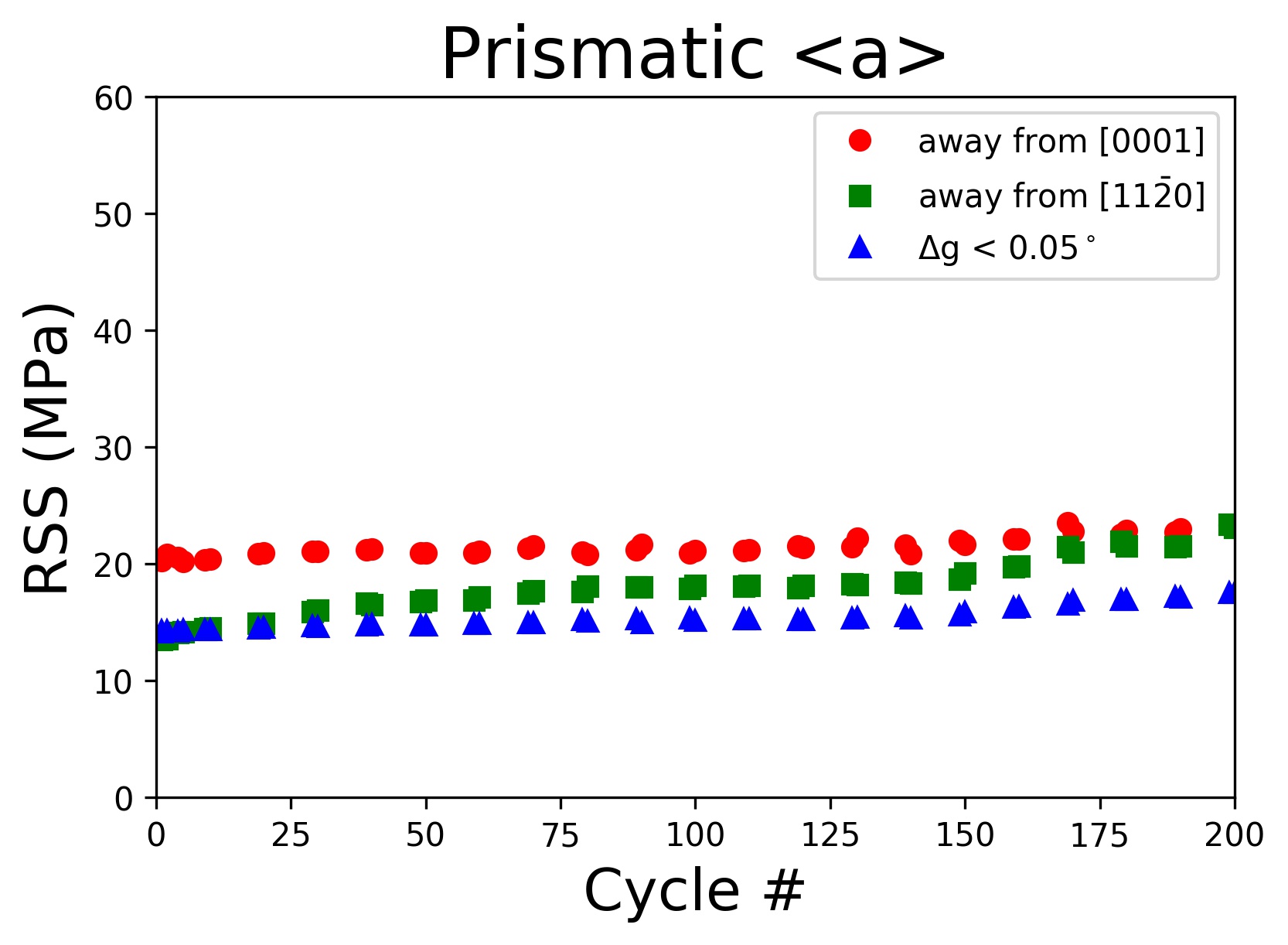}} \hspace{5pt}
		\subfloat{\includegraphics[width=0.48\textwidth]{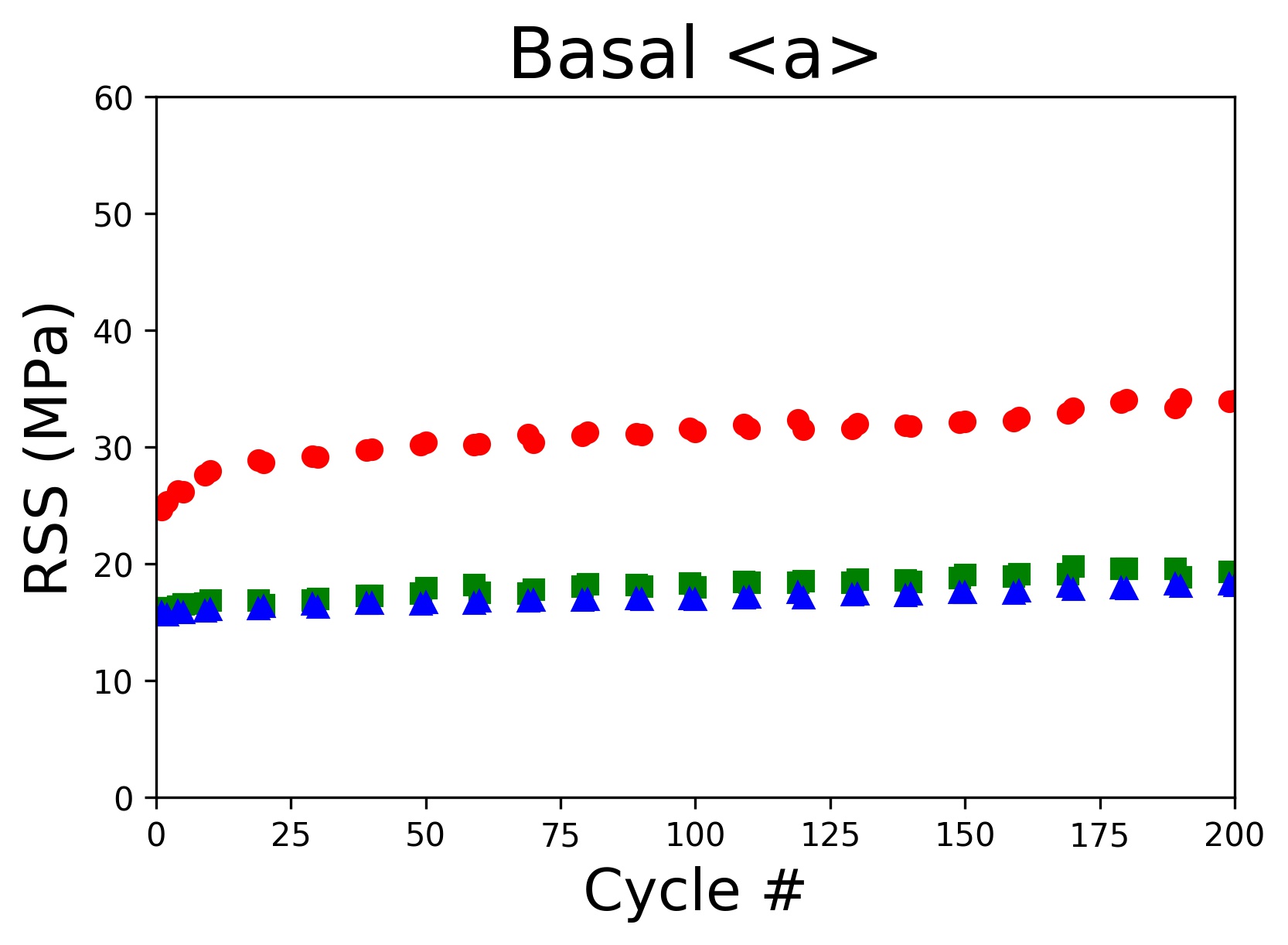}} \\
		\subfloat{\includegraphics[width=0.48\textwidth]{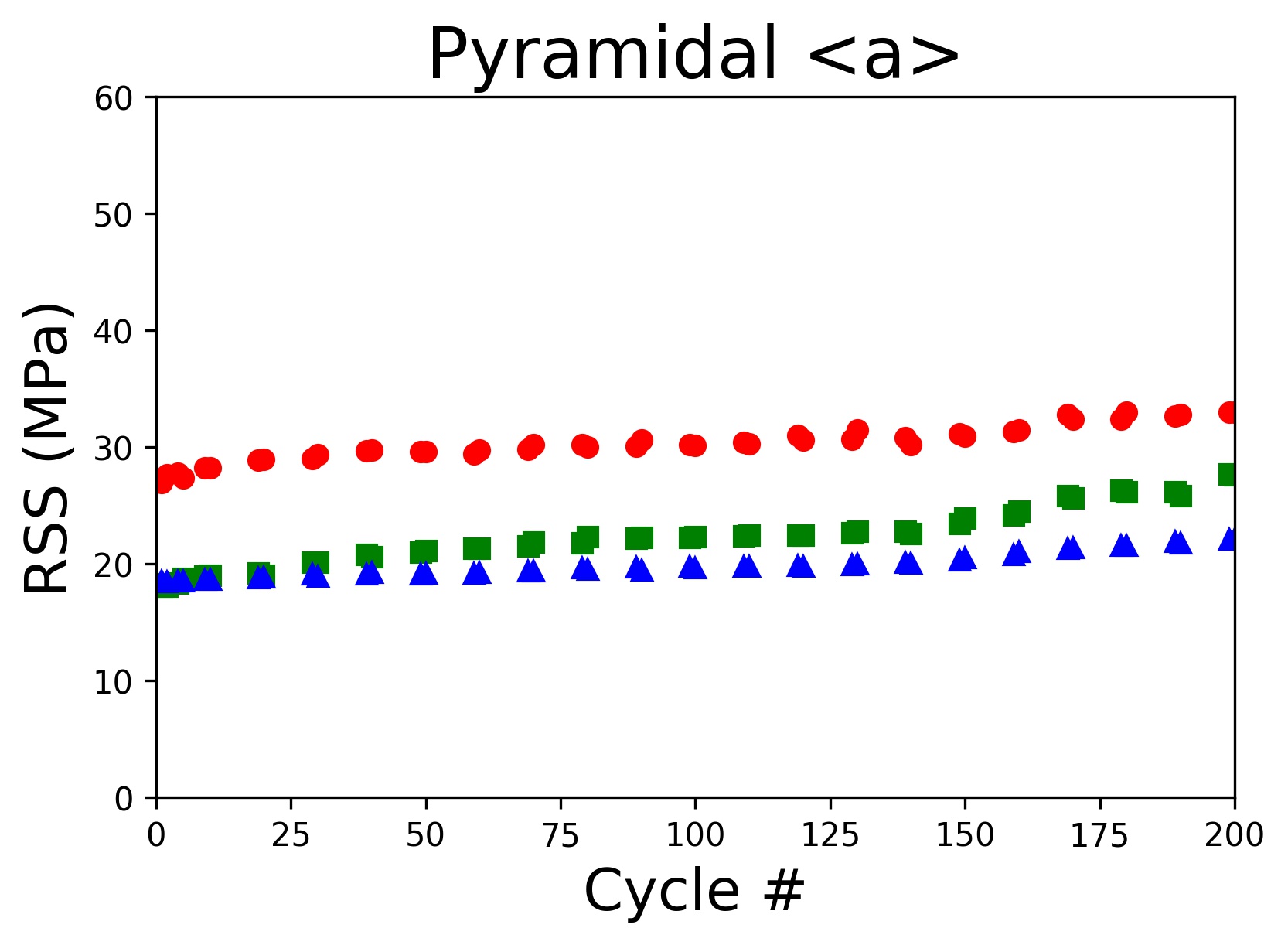}} \hspace{5pt}
		\subfloat{\includegraphics[width=0.48\textwidth]{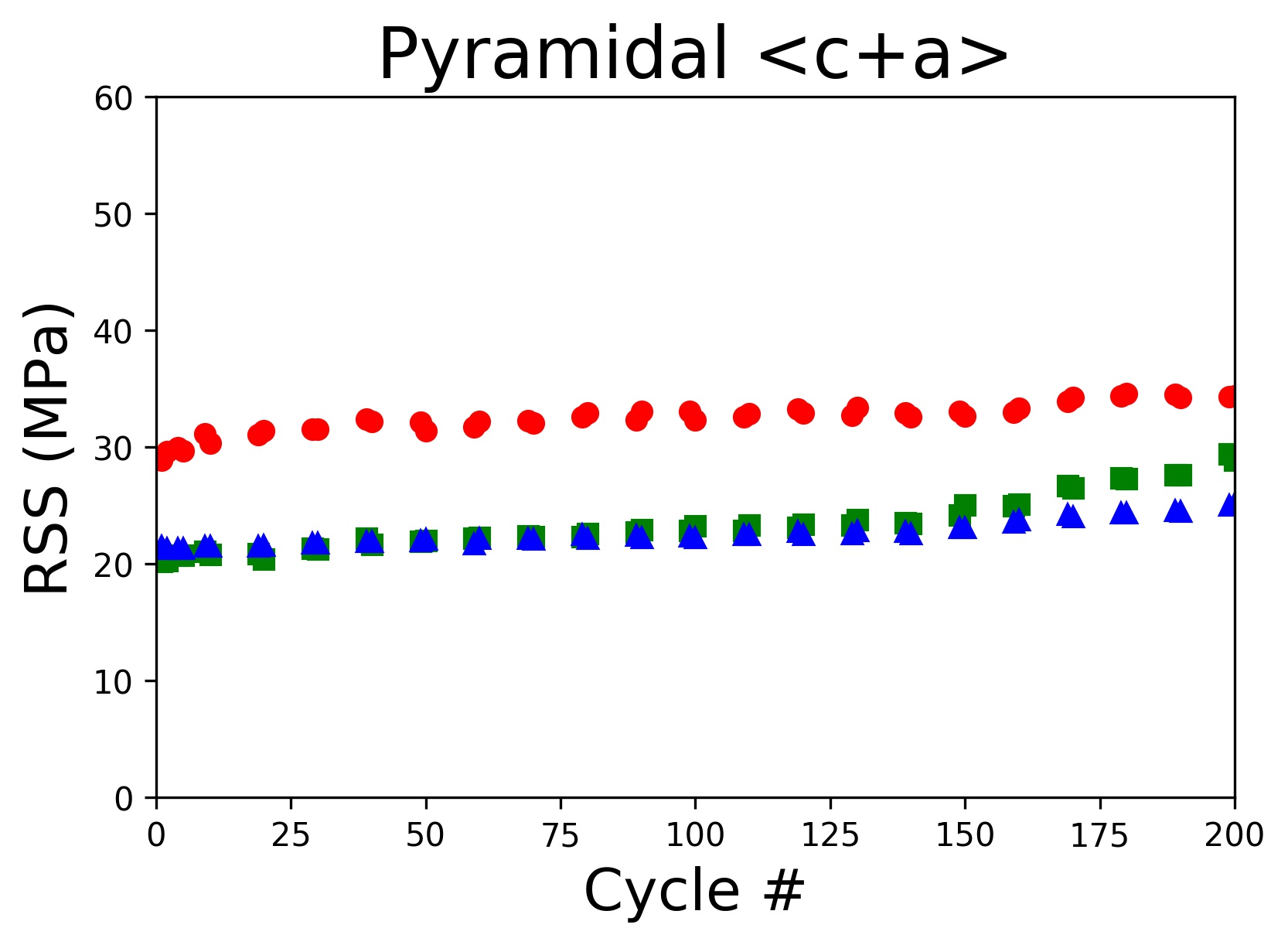}} \\
		\caption{In the unloaded state, the maximum RSS values for each slip system for each grain were averaged and plotted for each of the three categories of observed rotation behavior. The grains with the (0001) direction rotating away from the loading direction have a higher RSS on all of the slip systems than the rest of the grains.}
		\label{fig:RSS_noload}
	\end{figure}

	The collection of grains was divided into three dif{}ferent categories according to the direction of their grain rotation: grains with greater than 0.05$\degree$ change in rotation which are rotating away from [0001], grains with greater than 0.05$\degree$ change in rotation which are rotating away from [2$\bar{1}$$\bar{1}$0], and grains with less than 0.05$\degree$ change in rotation. The resolved shear stress (RSS) on each of the slip systems was calculated from the grain averaged stress tensors and the maximum value was found for each grain for each family of slip systems. Then, the average of these maximum values was found for each of the groups of grains and plotted with respect to cycle number. In the case of the unloaded state in Figure \ref{fig:RSS_noload}, the grains with their c-axis pointing close to the loading direction have higher RSS values than any of the other categories of grains. These grains are in a ``hard" orientation, which means they are more resistant to deformation. As such, the higher RSS values for these grains indicate that they are accumulating more of the residual stress as compared to the rest of the grains when the sample is unloaded.

\section{Summary}
	A sample of Ti-7Al was cycled 200 times at 90~\% of its yield stress, and f{}f-HEDM scans were taken in both the unloaded and loaded states for the 1st, 2nd, 5th, and 10th cycles, and every 10 cycles subsequently (Fig.~\ref{fig:ti7_stress_strain}). EDM of the sample left residual strain that is evident in the initial state (Fig.~\ref{fig:strain22_probplot}). However, we were able to resolve this residual strain and observed the high stress grains yielded during the first cycle. The evolution of the grains through the 200 cycles reveals a continual change in von Mises stress and orientation across the entire specimen (Figs.~\ref{fig:VM_probplot} and \ref{fig:ori_probplot}) indicating a slow accumulation of plasticity even though the sample was cycled below its macroscopic yield. In the loaded state, the grain stress tensors become progressively more aligned with the macroscopic stress tensor through increasing cycles (Fig. \ref{fig:cycles_loaded2}). Likewise, the grain stress tensors in the unloaded states become progressively less aligned with their initial grain stress tensor after the initial cycle (Fig. \ref{fig:cycles_loaded2}b). The six grains with the largest change in orientation through the 200 cycles all rotate such that their c-axes point further away from the loading direction (Fig.~\ref{fig:ipfs}). In general, the grains oriented with their c-axis more parallel to the loading direction slip on the basal planes and rotate away from the [0001] direction, while the grains closer to $[2\bar{1}\bar{1}0]$ slip on the prismatic system and rotate away from the $[11\bar{2}0]$ direction (Fig. \ref{fig:ipf_arrows}). This, in fact, matches predictions that can be made based on Schmid factor and CRSS adjusted for hardening behavior and slip rate dependence. Overall, the pattern of grain reorientation is consistent with the development of a tension texture despite the negligible magnitude of (macroscopic) plastic strain. It was observed that the grains in a ``hard" orientation (c-axis closer to the loading direction) retain more residual stress on all of the slip systems upon unloading than the grains in a ``soft" orientation (Fig. \ref{fig:RSS_noload}). Lastly, the precision of grain-averaged orientation measurements from the f{}f-HEDM technique permitted for the first time analysis of the upper tails of distributions of grain reorientation, which were fit well by the Fr\'{e}chet distribution: this quantification may provide input for the quantification of fatigue crack initiation, see, e.g. Przybyla \textit{et al.} \cite{Przybyla2012}.

\section*{Acknowledgements}
	The authors thank Adam Pilchak for providing the Ti-7Al material. They are also grateful to Christopher Kantzos, He Liu, and Yu-Feng Shen for helping with the experiment. Joe Pauza, Vahid Tari, Sven Gustafson, Jerard Gordon, and Bill Musinski are thanked for helpful discussions. This work is based upon research conducted at the Cornell High Energy Synchrotron Source (CHESS) which was supported by the National Science Foundation and the National Institutes of Health/National Institute of General Medical Sciences under NSF award DMR-1332208. The work is funded by the Air Force Of{}fice of Scientific Research under grant FA9550-16-1-0105.

\newpage
\clearpage

\bibliographystyle{elsarticle-num}
\bibliography{References}

\end{document}